\documentclass[pdflatex,sn-basic]{sn-jnl}

\usepackage{graphicx}%
\usepackage{multirow}%
\usepackage{amsmath,amssymb,amsfonts}%
\usepackage{amsthm}%
\usepackage{mathrsfs}%
\usepackage[title]{appendix}%
\usepackage{xcolor}%
\usepackage{textcomp}%
\usepackage{manyfoot}%
\usepackage{booktabs}%
\usepackage{algorithm}%
\usepackage{algorithmicx}%
\usepackage{algpseudocode}%
\usepackage{listings}%
\usepackage{upgreek}
\usepackage{aas_macros} 
\makeatletter
\def\ref@jnl#1{{\jnl@style#1\ }}
\makeatother
\setcitestyle{aysep={}}

\renewcommand{\mu}{\upmu}

\begin{document}

\title[Experimental studies of black holes: status and future prospects]{Experimental studies of black holes: status and future prospects}


\author*[1,2,3]{\fnm{Reinhard} \sur{Genzel}}\email{genzel@mpe.mpg.de}

\author[1,4]{\fnm{Frank} \sur{Eisenhauer}}\email{eisenhau@mpe.mpg.de}

\author[1]{\fnm{Stefan} \sur{Gillessen}}\email{ste@mpe.mpg.de}

\affil*[1]{\orgname{Max-Planck Institute for Extraterrestrial Physics}, \orgaddress{\street{Gie{\ss}enbachstr. 1}, \postcode{85748} \city{Garching}, \country{Germany}}}

\affil[2]{\orgdiv{Departments of Physics \& Astronomy}, \orgname{University of California}, \orgaddress{\street{Le Conte Hall}, \city{Berkeley}, \state{CA}, \postcode{94720}, \country{USA}}}

\affil[3]{\orgdiv{Faculty of Physics}, \orgname{Ludwig Maximilian University},  \orgaddress{\city{Munich}, \country{Germany}}}

\affil[4]{\orgdiv{Department of Physics}, \orgname{TUM School of Natural Sciences, Technical University of Munich}, \orgaddress{\postcode{85748} \city{Garching}, \country{Germany}}}


\abstract{More than a century ago, Albert Einstein presented his general theory
  of gravitation (GR) to the Prussian Academy of Sciences. One of the
  predictions of the theory is that not only particles and objects with
  mass, but also the quanta of light, photons, are tied to the curvature
  of space-time, and thus to gravity. There must be a critical compactness, above which photons cannot escape. These are black holes
  (henceforth BH). It took fifty years after the theory was announced
  before possible candidate objects were identified by observational
  astronomy. And another fifty years have passed, until we finally have
  in hand detailed and credible experimental evidence that BHs of 10 to
  $10^{10}$ times the mass of the Sun exist in the
  Universe. Three very different experimental techniques, but all based
  on Michelson interferometry or Fourier-inversion spatial
  interferometry have enabled the critical experimental breakthroughs.
  It has now become possible to investigate the space-time structure in
  the vicinity of the event horizons of BHs. We briefly summarize these
  interferometric techniques, and discuss the spectacular recent
  improvements achieved with all three techniques. Finally, we sketch
  where the path of exploration and inquiry may go on in the next
  decades.}


\maketitle

\setcounter{tocdepth}{3} 
\tableofcontents

\section{\textit{Presto:} Theoretical background}\label{sec:1}

A `\emph{black hole}' (e.g. \citealt{wheeler1968}) conceptually is a
region of space-time where gravity is so strong that within its
\emph{event horizon} neither particles with mass, nor even
electromagnetic radiation, can escape from it. Based on Newton's theory
of gravity and assuming a corpuscular nature of light, Rev.\ John Michell
(in \citeyear{michell1784}) and Pierre-Simon Laplace (in \citeyear{laplace1795}) were the first to note that
a sufficiently compact, massive star may have a surface escape velocity
exceeding the speed of light. Such an object would thus be `dark' or
invisible. A proper mathematical treatment of this remarkable
proposition had to await Albert Einstein's theory of General Relativity
in 1915/1916 (\citealt{einstein1916}, henceforth GR). Karl Schwarzschild's (\citeyear{schwarzschild1916})
first analytic solution of the vacuum field equations in spherical
symmetry revealed the unavoidable existence of a characteristic
\emph{event horizon} in the \emph{metric} of a mass M,
the \emph{Schwarzschild radius}
$R_{s}=2\,GM/c^{2}= 2\,R_{g}$
(with the gravitational radius $R_{g} =
GM/c^{2}$), within which no communication is possible
with external observers. It is a `one way door'. Radially inward moving
observers, after crossing the event horizon, cannot stop, nor reverse
back out, but end up in finite `Eigenzeit' (proper time) at the center.
All the mass/energy of a BH is concentrated there in a central
singularity.

\cite{kerr1963} generalized this solution to spinning BHs. For the
normalized spin parameter $(0\leq \chi \leq 1)$ the event horizon becomes
\begin{equation}
R_{\rm event\ horizon} = \frac{GM}{c^{2}} \times \left( 1 + (1 - \chi^{2})^{1/2} \right).
\label{eq:1}
\end{equation}

In \citeyear{newman1965}, Newman found the {axisymmetric} solution
for a BH that is both rotating and {electrically
charged}. Israel, Carter, Robinson, Wheeler, Bekenstein and Ruffini then
formulated the so-called
`\emph{no-hair theorem}' (1967\,--\,1975)\footnote{See \cite{israel1967, carter1971,
  robinson1975} and \cite{bekenstein1975}}, stating that \emph{a
stationary BH solution is completely described by the three parameters
of the {Kerr--Newman metric}: {mass}, {angular momentum}, and electric charge}. For the Kerr metric this means that the
quadrupole moment $Q_{2}$ of the BH is determined
by the spin, namely $Q_{2}/M =-\chi^{2}$.
However, these solutions refer to configurations with sufficiently high
symmetry, so that Einstein's equations can be solved
analytically. This led to a debate whether the conclusions obtained were
generally applicable. \cite{penrose1963,penrose1965} dropped the assumption of
spherical symmetry, and analyzed the problem topologically. Using the
key concept of `\emph{trapped surfaces}' he showed that any
arbitrarily shaped surface with a curvature radius less than the
Schwarzschild radius is a trapped surface. Any observer is then
inexorably pulled towards the center where time ends.

The distortion of the space-time outside the event horizon leads to a
minimum radius, where stable circular orbits are possible. For particles
with mass this \emph{innermost stable, circular orbital radius}
(ISCO) is $6\, R_{g}=3\, R_{S}$ for
$\chi=0$, and $R_{g}$ for $\chi=1$. For photons of
no mass this innermost stable orbital radius (called
\emph{photon orbit}) is $3\, R_{g}=1.5$
$R_{S}$ for $\chi=0$, and $R_{g}$ for
$\chi=1$ \citep{bardeen1972}. Finally, if a BH is irradiated by a
point source at large distance behind the BH, only photons with
projected radii $\geq 3 \sqrt{3} R_{g}$ arrive at the distant
observer in front of the BH. Those inside form a
`\emph{{shadow}}' (a central depression of light, \citealt{bardeen1972, luminet1979, falcke2000a}) and do not reach the
observer.

Work by Bardeen, Bekenstein, Carter, Christodoulou, Ruffini and Hawking in the early
1970s\footnote{See \cite{christodoulou1970, christodoulou1971, carter1971, bardeen1973a, bardeen1973b, hawking1974,
  bekenstein1975}} led to the formulation of
{\emph{BH thermodynamics}}. These laws describe the behavior of a BH in close
analogy to the {laws of classical thermodynamics}, by relating mass to energy, area to {entropy}, and {surface gravity} to {temperature}. The analogy was completed when \cite{hawking1974} showed that {quantum field theory} implies that BHs should emit particles and photons like a {black body} with a temperature proportional to the surface gravity of the BH, hence
inversely proportionally to its mass. This predicted effect is now known
as {\emph{Hawking radiation}}'. For the astrophysical BHs discussed here, the Hawking radiation is out of reach of current detection methods by many orders of
magnitude.

From considerations of the \emph{information content} of BHs,
there is significant tension between the predictions of GR and general
concepts of quantum theory (e.g., \citealt{susskind1995, maldacena1998, bousso2002}). It is likely that a proper quantum theory of gravity will modify
the concepts of GR on scales comparable to or smaller than the
\emph{Planck length}, 
$l_{\rm Pl}=\sqrt{\frac{\hslash G}{c^3}}\sim1.6\times10^{-33}\, \mathrm{cm}$, remove the concept of the central singularity, and potentially
challenge the interpretation of the GR event horizon \citep{almheiri2013}. If gravity is fundamentally a higher-dimensional interaction,
then the fundamental Planck length in 3D can be substantially larger
\citep*{arkani-hamed1998}.

But are these bizarre objects of GR (and science fiction) actually realized in Nature? The
ultimate question discussed in the following is not just whether the
weak-ﬁeld gravity region near compact objects is qualitatively
consistent with the BH geometry of GR, but rather to quantify the limits
of observations (= experiments at a distance) in testing the existence of event horizons (cf.\ \citealt{cardoso2019}, and references therein). ``How close''
is a self-gravitating object to a BH? One can introduce a
\emph{``closeness'' parameter $\epsilon$, such that $\epsilon \to  0$ corresponds to
the BH limit}. For example one can choose the compactness, such that
for a spherically symmetric space time
\begin{equation}
    \epsilon = 1 - \frac{2 \, GM/c^2}{R} \,
    \label{eq:2}
\end{equation}
where $M$ is the object mass in the static case and $R$ is its
radius (cf.\ \citealt{cardoso2019}). Likewise, one can introduce $\epsilon$ as a measure of
the violation of the no hair theorem above. Alternatives of the GR BH
proposal for compact astrophysical objects are `\emph{exotic
compact objects}' ('ECOs', \citealt{cardoso2019,psaltis2023}). These might be concentrations of heavy,
dark matter bosons or fermions, such as `\emph{{boson stars}}'
(Torres et al. 2000), or `\emph{fermion balls}' \citep{viollier1993, tsiklauri1998, becerra-vergara2020}, or
`\emph{grava-stars}' (stars supported by negative vacuum
pressure, e.g., \citealt{mazur2004, cardoso2019}), or
`\emph{worm-holes'} \citep{morris1988, cardoso2019}.

\section{\textit{Vivace:} X-ray binaries \& quasars}\label{sec:2}

Astronomical evidence for the existence of BHs started to emerge sixty
years ago with the discovery of variable X-ray emitting binary stars in
the Milky Way \citep{giacconi1962, giacconi2003} on the one hand, and
of distant, luminous `quasi-stellar-radio-sources/objects'
(\emph{{quasars}} or \emph{{QSO}}s, \citealt{schmidt1963}) on the
other. Dynamical mass determinations from Doppler spectroscopy of the
visible primary star established that the mass of the X-ray emitting
secondary is sometimes significantly larger than the maximum stable neutron star
mass, $\sim$2.3 solar masses \citep{mcclintock2004,
remillard2006, ozel2010, rezzolla2018}.
The binary X-ray sources thus are excellent candidates for stellar BHs
(SBH, $\sim8$--$20\, M_{\odot}$). If so they are probably
formed when a massive star explodes as a supernova at the end of its
fusion lifetime and the compact remnant collapses to an SBH.

The radio to X-ray luminosities of quasars often exceed by 3 to 4 orders
of magnitude the entire energy output of the Milky Way Galaxy.
Furthermore, their strong high-energy emission in the UV-, X-ray and
$\gamma$-ray bands, as well as their spectacular relativistic jets, can most
plausibly be explained by accretion of matter onto rotating (super)-
massive BHs (henceforth (S)MBHs,
$10^{6}$--$10^{10}\, M_{\odot}$, e.g.,
\citealt{lynden-bell1969, shakura1973, blandford1977,
rees1984, blandford1999b, yuan2014, blandford2019}). Between 5.7\% (for a non-rotating Schwarzschild hole) and
42\% (for a maximally rotating Kerr hole) of the rest energy of
infalling matter can, in principle, be converted to radiation outside
the event horizon. This efficiency is two orders of magnitude greater
than nuclear fusion in stars. To explain powerful QSOs by this
mechanism, BH masses of $10^{8}$ to $10^{10}$
solar masses, and accretion flows between 0.1 to tens of solar masses
per year are required. Often the accretion rate is expressed as \textit{Eddington ratio}, where a value of 1 corresponds to the situation that the radiation pressure of the emission equals the gravitational pull of the MBH.

Quasars are located (without exception) in the nuclei of large, massive
galaxies (e.g., \citealt{osmer2004}). Quasars represent the most extreme and
spectacular among the general nuclear activity of most galaxies \citep{netzer2015}. There may also be intermediate mass BHs (IMBHs,
$10^{2}$--$10^{5}\, M_{\odot}$), for
instance in the cores of globular clusters or dwarf galaxies. Evidence
for $>10^{5}\,M_{\odot}$ MBHs in low mass
galaxies is growing but the case for IMBHs in globular clusters is still
very much debated \citep{greene2020}. Finally, there have been
proposals that BHs with a wide mass spectrum might have been created in
the rapid cool-down phase after the Big Bang (e.g., \citealt{carr1974,
carr1975, hasinger2020}).

A conclusive experimental proof of the existence of a BH, as
defined by GR, requires the \emph{determination of the
gravitational potential} \emph{at or near the scale of the
event horizon}. This gravitational potential can be inferred from
spatially resolved measurements of the motions of test particles
(interstellar gas, stars, other BHs, or photons) in close trajectory
around the BH \citep{lynden-bell1971}, or from gravitational waves
emitted in the inspiral of a binary BH. \cite{lynden-bell1969} and
\cite{lynden-bell1971} proposed that (S)MBHs might be common in most
galaxies (although in a low state of accretion). If so, dynamical tests
are feasible in nearby galaxy nuclei, including the center of our Milky
Way. Because of the small angular radius of the event horizon (e.g. 10
micro-arcsec for the 4.3 million solar mass MBH even in the `nearby',
8.27 kpc, Galactic Center), achieving the necessary instrumental
resolution requires \emph{extremely large telescopes (or spatial
interferometers) with exquisite sensitivity and spectral resolution}.

Over the past fifty years, increasingly solid \emph{evidence for
central `dark' (i.e. non-stellar) mass concentrations} has emerged for
more than several hundred galaxies in the local Universe (e.g., \citealt{magorrian1998, gebhardt2000, ferrarese2000, kormendy2004a, gultekin2009, fabian2012, kormendy2013, mcconnell2013, saglia2016, greene2013, greene2016}). The data come
from optical/infrared imaging and spectroscopy on the Hubble Space
Telescope (HST, and most recently from the James Webb Telescope (JWST)),
from large ground-based telescopes, as well as from Very Long Baseline
radio Interferometry (VLBI). Further evidence comes from
relativistically broadened, redshifted iron K$\alpha$ line emission in nearby
Seyfert galaxies (e.g. \citealt{tanaka1995, nandra1997, fabian2000}), including the first statistical constraints on the BH
spin distribution \citep{reynolds2021}.

In external galaxies a compelling case that such a dark mass
concentration cannot just be a dense nuclear cluster of white dwarfs,
neutron stars and perhaps stellar BHs, already emerged in the mid-1990s
from spectacular VLBI observations of the nucleus of NGC 4258. This is a
mildly active galaxy at a distance of 7~Mpc \citep{miyoshi1995, moran2008}. The VLBI observations show that the galaxy nucleus contains a
thin, slightly warped disk of H\textsubscript{2}O masers (viewed almost
edge on), in beautiful Keplerian rotation around an unresolved mass of
40~million solar masses, as inferred from the maser motions. The maser
motions exceed 1000 km/s at the innermost edge of the disk of about
0.1pc. The inferred density of this mass exceeds a few
$10^{9}$ solar masses pc$^{-3}$ and thus
cannot be a long-lived cluster of `dark' astrophysical objects of the
type mentioned above \citep{maoz1995}. \cite{greene2013} presented a
survey of such H\textsubscript{2}O disk maser MBHs. As we will discuss
below, the Galactic Center provides a yet more compelling case.

\begin{figure}[ht]
    \centering
    \includegraphics[width=1\linewidth]{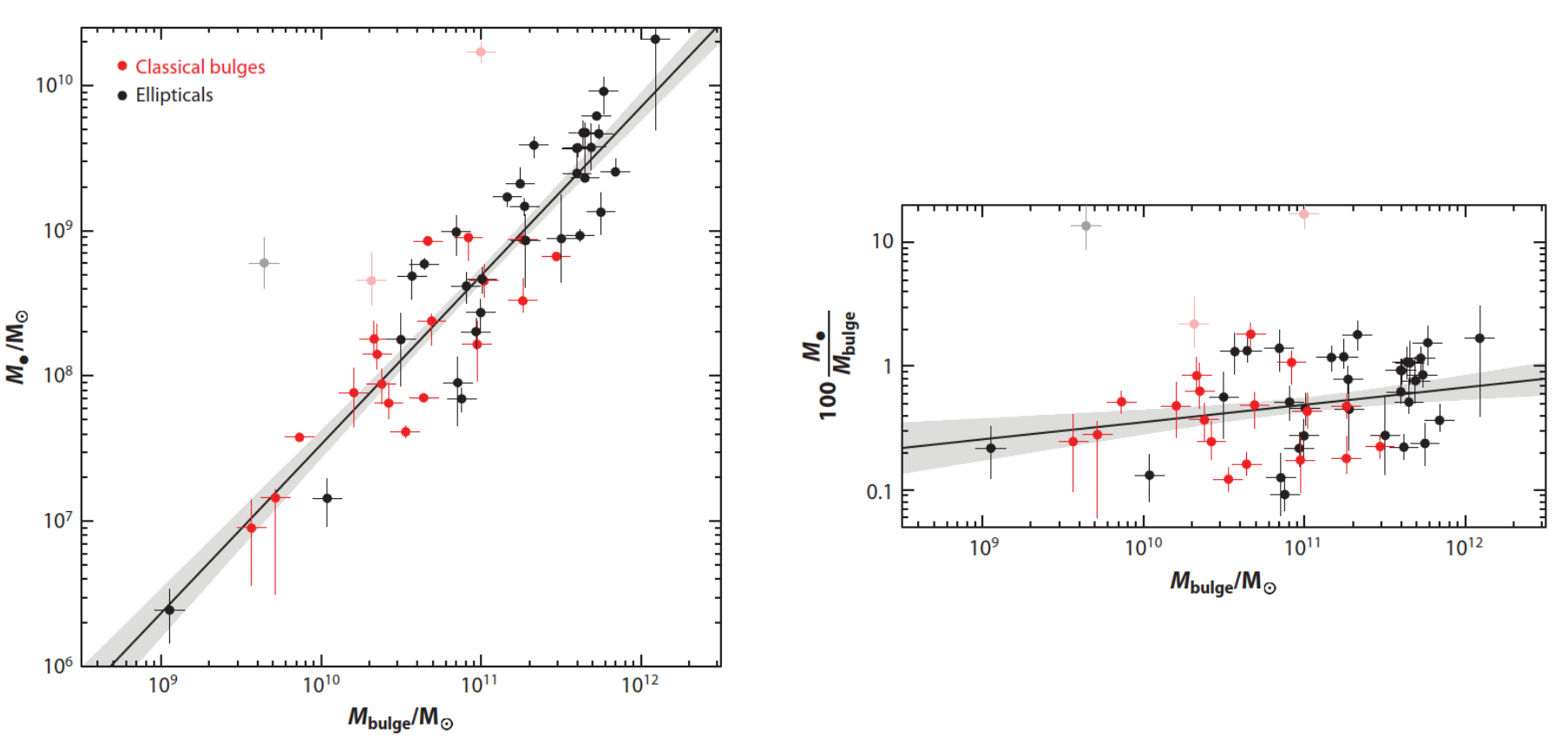}
    \caption{Relationship between inferred central mass (from stellar and
gas dynamics) $M_{\bullet}$ and bulge mass in local Universe
massive galaxies (ellipticals, or disks with classical bulges). There
clearly is a correlation between these two components, with a best fit
$\frac{M_{\bullet}}{M_{\rm bulge}}=(4.9 \pm 0.6)\times 10^{-3}\times \left( \frac{M_{\rm bulge}}{10^{11}\,M_{\odot}}\right)^{0.14 \pm 0.08}$ (adapted
from Fig.~18 of \citealt{kormendy2013}).}
    \label{fig:1}
\end{figure}

In the galaxies investigated, dark masses are found ranging from a few
$10^{4}$ to $10^{5}\, M_{\odot}$ in low
mass systems \citep{greene2020}, to
$10^{10+}\,M_{\odot}$ in very massive
spheroidal/elliptical galaxies \citep{kormendy2013, mcconnell2013}. For the ellipticals and for galaxies with `classical' bulges
\citep{kormendy2004b}, there appears to be a fairly low scatter
relationship between central mass and bulge mass \citep{haring2004,
kormendy2013, mcconnell2013}. About 0.2--0.7\% of the bulge
mass is in the central dark mass, increasing slowly with bulge
mass, and strongly suggesting that central dark mass and bulge have
grown together over cosmological time scales (Fig.~\ref{fig:1}).

\section{\textit{Allegro:} Testing the MBH paradigm in the Galactic Center
with stellar orbits and radio
emission}\label{sec:3}

\begin{figure}[htbp!]
    \centering
    \includegraphics[width=1\linewidth]{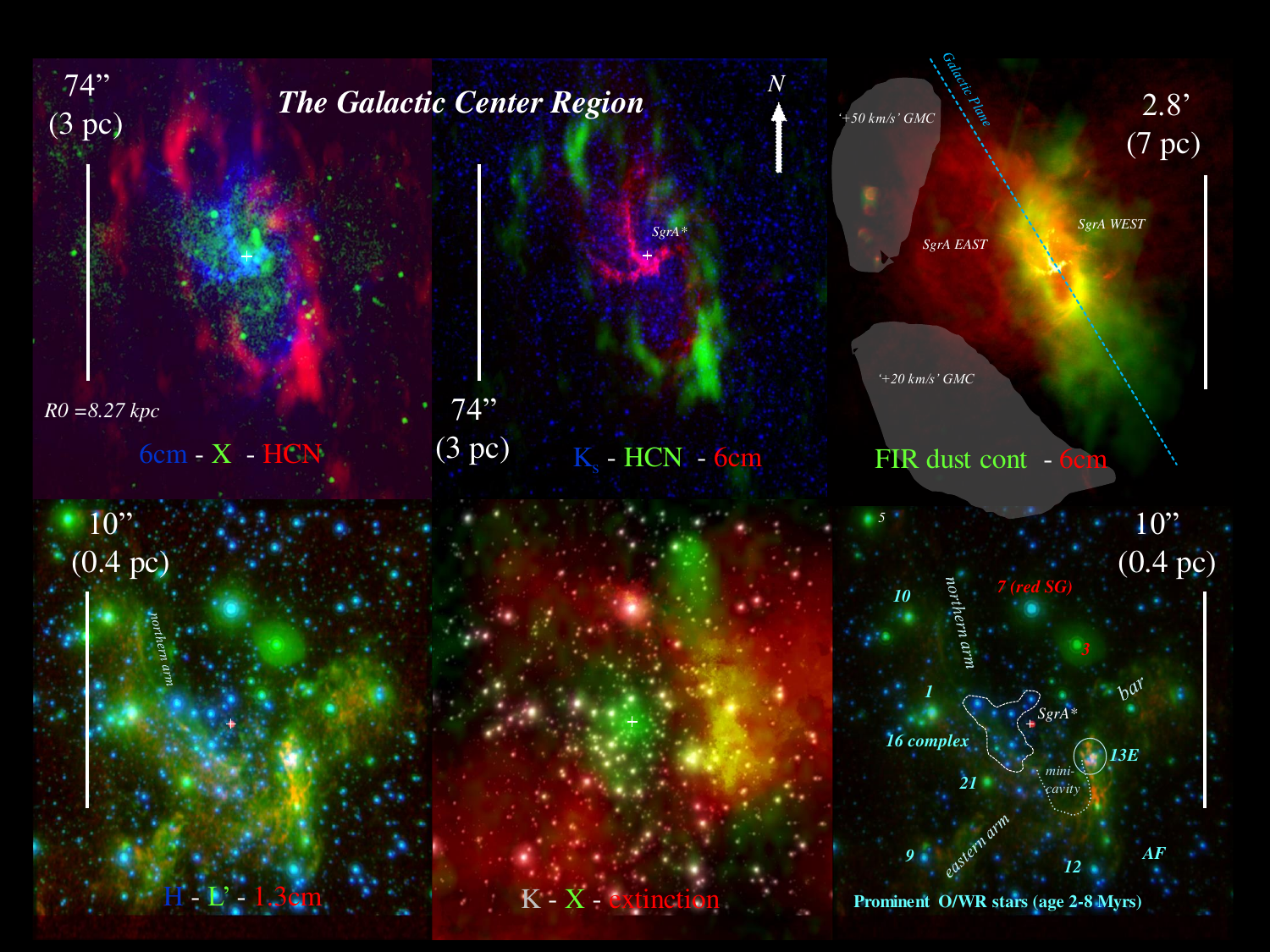}
    \caption{Summary of the different components: stars (old giants, red
and blue super-giants), cold (20--200 K) molecular/neutral gas and dust,
ionized ($10^{4}$ K) and hot (1--$10\times10^{6}$ K)
gas, and their distributions on sub-parsec, to 10 parsec scale in the
Galactic Center (adapted from \citealt{genzel2010}). The cross in the center of the
images marks the location of the compact, non-thermal radio source
SgrA*, probably a MBH of 4.3 million solar masses \citep{genzel2010}. \textbf{Top right:}
Largest scales of the SgrA region, with the HII region SgrA WEST and the
supernova remnant SgrA EAST (presumably an explosion of one or several
massive O/Wolf--Rayet star(s) $\sim$20--40,000 years ago. Outside
of this region are two giant molecular clouds at `+20' and `+50' km/s
LSR velocity. \textbf{Top left and center:} zoom in onto SgrA WEST, which harbors
the center of a dense
($\rho_{*} > 10^{6}\, M_{\odot}\, \mathrm{pc}^{-3}$) cluster of old, and young,
massive stars. The central 1.5-pc diameter region is filled with ionized
gas streamers (bottom left), hot X-ray emitting gas (bottom center), and
the most massive, recently formed O, Wolf--Rayet and B-stars (bottom
right). Winds and UV-radiation from these stars and the MBH have created
a lower density `cavity' relatively devoid of dense molecular gas and
dust (average hydrogen density
$n_{H}\sim10^{3..4.5}\, \mathrm{cm}^{-3}$). The central cavity in turn is surrounded by a
rotating, clumpy `circum-nuclear' ring of warm dust and dense, molecular
gas (HCN and other high excitation gas components are found here, and
the molecular hydrogen density is
$n_{\rm H2}\sim10^{5-6}\, \mathrm{cm}^{-3}$, \citealt{becklin1982, ho1995}). Gas is
streaming in and out of the central region in form of clumpy, tidally
disrupted `streamers', such as the `northern' and `eastern' arms and the
`bar' (cf.\ \citealt{oort1977, lo1983, genzel1987, ho1991, genzel1994, melia2001, genzel2010, morris2012}).}
    \label{fig:2}
\end{figure}

The central light years of our Galaxy contain a dense and luminous star
cluster, as well as several components of neutral, ionized and extremely
hot gas (Fig.~\ref{fig:2}; \citealt{genzel1987, genzel1994, morris1996, melia2001, genzel2010, morris2012}).
Compared to the distant quasars, the Galactic Center is `just around the
corner' ($R_0=8.27$ kilo-parsecs (kpc), 27,000 light
years). High resolution observations of the Milky Way nucleus thus offer
the unique opportunity of carrying out a stringent test of the
MBH-paradigm deep within its gravitational `sphere of influence' where
gravity is dominated by the central mass ($R<1-3$ pc). Since the
center of the Milky Way is highly obscured by interstellar dust
particles in the plane of the Galactic disk, observations in the visible
part of the electromagnetic spectrum are not possible. The veil of dust,
however, becomes transparent at longer wavelengths (the infrared,
microwave and radio bands), as well as at shorter wavelengths (hard
X-ray and $\gamma$-ray bands), where observations of the Galactic Center thus
become feasible \citep{oort1977}.

\subsection{Initial statistical evidence for a compact central mass
from gas and stellar motions}

Starting in the late 1970s/1980s, observations of the Doppler motions of
ionized and neutral gas clouds in the central parsecs \citep{wollman1977, lacy1980, serabyn1985, crawford1985}, and
of stellar velocities \citep{mcginn1989, krabbe1995, haller1996} found the first evidence for a central mass concentration of a
few million solar masses, concentrated on or near the compact radio
source SgrA*. In the 1990s observations of stellar proper motions with
the telescopes of the European Southern Observatory (ESO) in Chile \citep{eckart1996, genzel1997}, and with the Keck
telescopes on Mauna Kea \citep{ghez1998} further improved the
statistical and systematic evidence. Yet in terms of the compactness parameter $\epsilon$
introduced in the first paragraph, these early measurements did not
provide significant evidence that this mass concentration must be a BH:
$0 \ll \epsilon \sim 1$--$10^{-5}$. It could instead be a cluster of faint stars,
neutron stars, or stellar BHs.

\subsection{Sharper images and individual stellar orbits on solar
system scales}

Further progress required three new key elements. One is much higher
angular resolution and integral-field imaging spectroscopy (achieved
with 8--10-m telescopes, and aided by adaptive optics to reach the
diffraction limit of $\sim$50--60 milli-arcsec). These
improvements were realized both in Chile (ESO-VLT) and in Hawaii (Keck)
between 2000 and 2005. The second are very long duration observation
campaigns ($>$1--2 decades) to observe not only stellar
velocities, but derive the full orbital parameters of individual stars
as precision tracers of the potential. The third was luck, namely to
find stars much closer to SgrA* than was theoretically expected (i.e., on
solar system scales. cf.\ \citealt{alexander2005, alexander2017}).

The most important scientific breakthrough started in 2002 when both the
ESO-VLT \citep{schodel2002} and the Keck telescope \citep{ghez2003}
discovered that the star S2 (or S02 in UCLA nomenclature) approached
SgrA* to about 15 milli-arcsec ($\sim$15 light hours or 1200
$R_{S}$), and sharply turned around SgrA* on a highly
elliptical orbit ($e=0.88$). By 2010, both the VLT- and the Keck-based
groups were able to derive orbits for about 10--20 stars remarkably close
to SgrA* (top left panel of Fig.~\ref{fig:3a}, \citealt{ghez2008, gillessen2009}), followed in the next decade by steady progress in the number and stars and quality
of their derived orbital parameters \citep{boehle2016, gillessen2017}. Now the observations were able to exclude the compact star
cluster hypothesis, but a few speculative alternative explanations to a
MBH, such as `boson' or `fermion' stars (see Sects.~\ref{sec:1} and \ref{sec:6}) still were
possible. And in any case, there also remained the theoretical
possibility that GR was not applicable, since not yet tested in the MBH
regime.

\subsection{Interferometry and detection of post-Newtonian orbital
deviations}

The next big breakthrough started in 2016/2017 with the completion and
continuous further improvements of the GRAVITY \citep{gravity2017} and
GRAVITY+ \citep{gravityplus2022} infrared interferometric beam combiner of
the $4\times8$-m telescopes of the ESO-VLT
(\url{https://www.mpe.mpg.de/ir/gravity},
Fig.~\ref{fig:A1a} left panel, \citealt{gravity2022a,eisenhauer2023}). GRAVITY(+) achieves a near-infrared resolution of
about 3 milli-arcsec, and an astrometric precision of 10--100
micro-arcseconds, about 10--20 times better than with individual
10m-class telescopes, and has polarization, spectroscopic, and wide and
deep imaging modes \citep{gravity2017, gravity2022b, gravity2024, gravityplus2022, gravityplus2024}.

\begin{figure}[ht!]
    \centering
    \includegraphics[width=1\linewidth]{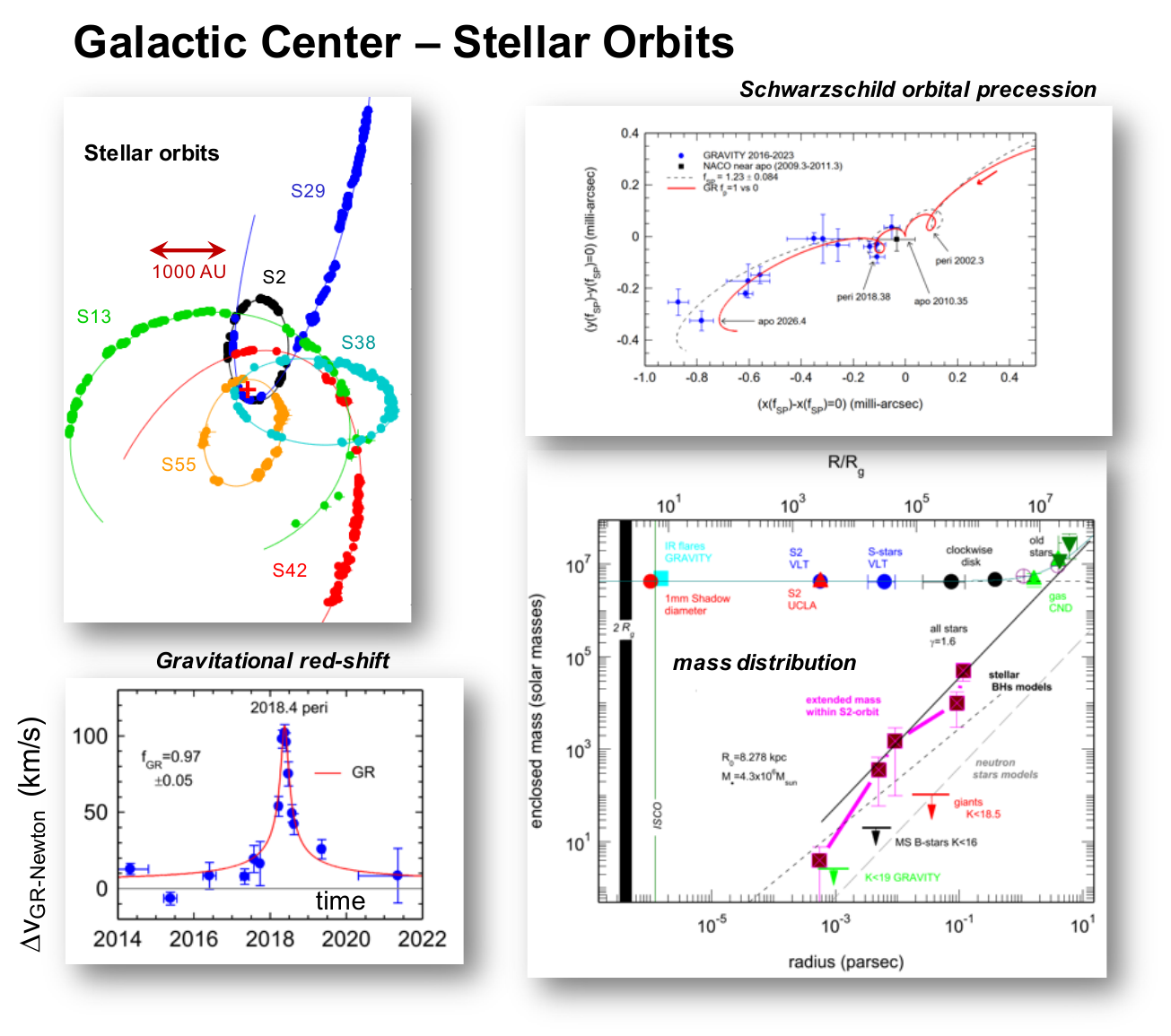}
    \caption{Tests of the MBH paradigm and GR in the Galactic Center using individual stellar orbits}
    \label{fig:3a}
\end{figure}

\begin{figure}[ht!]
    \centering
    \includegraphics[width=1\linewidth]{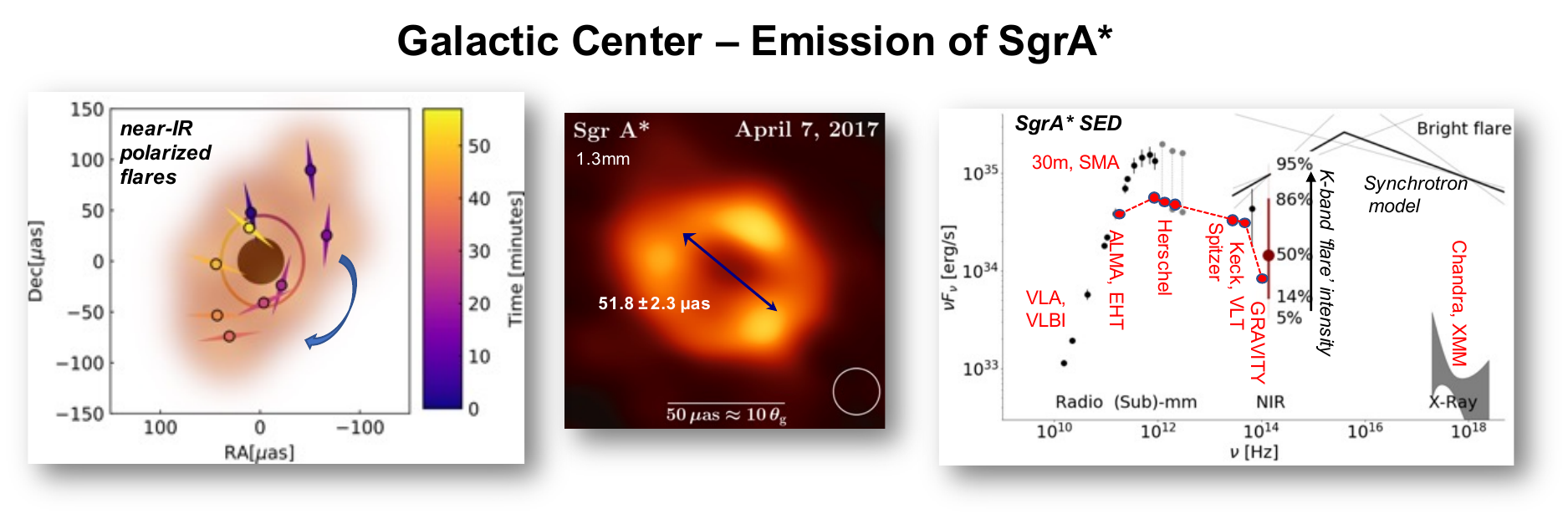}
    \caption{Evidence for the concentration of the mass of $4.3\times 10^6 M_\odot$ (as determined from stellar orbits) within few gravitational radii. Left: The flaring NIR emission of SgrA* revolves on a time scale compatible with the Keplerian period at the observed radius. The polarization of the NIR emission rotates on the same time scale. Middle: The EHT-image of SgrA* shows a silhouette as expected for an MBH of $4.3\times 10^6 M_\odot$. Right: The SED of SgrA*.}
    \label{fig:3b}
\end{figure}

With the combined new instrumental capabilities, precision measurements
of $\sim$50 stellar orbits (central panel) of the so-called
S-stars around SgrA* determine the central mass and its distance to the
Sun to be 4.300 ($\pm0.011$ ($1\sigma$ statistical), $\pm0.018$ ($1\sigma$ systematic))
$\times10^{6}\,M_{\odot}$, and $R_{0}=8273$
($\pm7.5$, $\pm15$) pc, centered on the position of SgrA* \citep{boehle2016,
gillessen2017, do2019a, gravity2019b, gravity2022a, gravity2024}. The stellar orbits set stringent limits to any additional
mass in the vicinity, of a few thousand solar masses within a few
$10^{4}\, R_{S}$ (bottom-right panel), as
determined by the orbital data of $\sim$15 stars \citep{gillessen2017, do2019a, gravity2019b, gravity2022a, gravity2022b, gravity2023a, evans2023, will2023}.

This conclusion is also supported the near-pericenter astrometric orbit of S2,
after subtracting the best fitting Newtonian orbit. While Newton's
theory would thus expect these positions to lie in an inclined highly
elliptical orbit centered on SgrA*, GR in contrast predicts that the
stellar motions exhibit a prograde in-plane precession, the
`\emph{Schwarzschild}'-precession, with an advance of
\[
\Delta\Phi \text{(per orbit)} = 3 \pi f_{\rm SP} \frac{R_{S}}{a(1-e^2)} \,.
\]
For the 16-year orbit of S2 ($a=0.125''$, or 1039 AU, $e=0.884$,
$R_{\rm peri}/R_{S}=a(1-e)/R_{S}=1200$) this precession is 12.1' per orbit in
clockwise direction. In the framework of post-Newtonian (PPN) expansions
\citep{johannsen2016a}, the Schwarzschild precession is first order in
$(v/c)^{2}$, or a few times $10^{-4}$.
GR has $f_{\rm SP}=1$, while Newton has
$f_{\rm GR}=0$. The top right panel of Fig.~\ref{fig:3a} shows the
location of the averaged S2 residual data obtained with GRAVITY (filled
blue circles) and NACO (VLT adaptive optics imaging, black cross), after
removing the Newtonian best-fit orbit. In this presentation the
precession predicted by GR turns into the loopy red line. The observed
data points are in excellent agreement with the GR prediction (the data
in Fig.~\ref{fig:3a} give $f_{\rm SP}=1.2\pm0.1$). This represents a
confirmation that GR is correct in the space-time environment of a 4
million solar mass concentration; it also confirms that this mass
concentration is compact and not a centrally peaked, but spatially
extended distribution \citep{gravity2020b,gravity2022a}. Most of the
mass is in a single concentrated object. Any nearby second (intermediate
mass) BH companion ($<10^{3}$ AU, or $10\, R_{\rm peri}$ of the stars S2 and S29), can only have a mass of $<10^{3}\, M_{\odot}$ \citep{gravity2023a, will2023}. In PPN=1.5
($\sim(v/c)^{3}$) GR has a second
orbital precession, the \emph{Lense--Thirring} precession per
orbit is 
\[
\Delta\Phi_{\rm LT} = 2 \chi \times \left(\frac{R_{S}}{a(1-e^2)}\right)^{1.5} \,,
\]
around the spin axis of the MBH. For the star S2, the Lense--Thirring
precession is about $0.0497\times \chi$ (arcminutes/orbit), for a spin
parameter $\chi (\leq 1)$, so at least 240 times smaller than the star's
Schwarzschild precession, and out of reach of the current astrometry.

GR also has a PPN=1 order effect in line of sight velocity, the
\emph{{gravitational redshift}}, of 100 km/s for S2. The bottom left
inset of Fig.~\ref{fig:3a} shows the residual Doppler velocity of the star S2 as
a function of time around the peri-approach in 2018.4, relative to that
predicted by a Newtonian orbit of the same orbital parameters
(horizontal grey line). The red curve are the residuals curve predicted
by GR. The actual observed residuals are the blue filled circles. Our
yield $f_{\rm GR}=0.97\pm0.05$ (in addition to another 100
km/s redshift due to the transverse Doppler effect), again in excellent
agreement with GR ($f_{\rm GR}=1.0$, \citealt{gravity2018a, gravity2020b, do2019a}).

\paragraph{Further tests of General Relativity near a MBH.}

\cite{gravity2019a} confirmed the Equivalence Principle in the
orbit of S2 through a test of the linear positional invariance. In that
paper the redshift data are split into spectroscopy of the hydrogen
Br$\gamma$ line and the HeI 2.1$\mu$m line, and the gravitational redshift term is
computed for the two data sets. Einstein's Equivalence Principle
stipulates that in free fall the motion should only depend on
mass/energy, and not on composition. And indeed, \cite{gravity2019a}  set an upper limit of a few $10^{-2}$ to the
fractional difference of the gravitational redshift in hydrogen and
helium. In another paper, \cite{hees2017, hees2020} used the Galactic
Center data to set limits on a hypothetical fifth force, and variations
in the fine structure constant. \citep{jovanovic2024} analyzed the
Schwarzschild precession of S2 in the framework of Yukawa gravity
theory, and set an upper limit to the mass of the graviton, which is
compatible with limits from aLIGO gravitational-wave data.

\paragraph{Near event horizon motions and strong magnetic fields.}

The near-IR emission from SgrA* itself is linearly polarized
($\sim$40\%, \citealt{eckart2006a, eckart2006b, genzel2010}) and is synchrotron
emission from very hot gas in the accretion zone, like the radio
emission. The near-IR emission is constantly varying, with a red-noise
power spectrum \citep{do2009,dodds2009,dodds2010,dodds2011,witzel2012,witzel2018}. The power spectrum is typically log-normal,
but occasionally high amplitude 'flares' above occur over a few hours at
$>$10--20 times the average quiescent level \citep{dodds2009, dodds2011, do2019b, genzel2010}. These flares
exhibit `clockwise' orbital motion on a scale of 8--9 $R_{g}$, just outside the EHT ring \citep{gravity2018b, gravity2023b}. The polarization direction also exhibits
rotation at the same rate and in the same direction as the astrometric
motions. \cite{gravity2023b} conclude that the accretion zone
must be within a few tens of degrees of face-on. The polarization
properties clearly show that that the near-event horizon accretion zone
is magnetically dominated with a dominant poloidal field (Fig.~\ref{fig:3b}:
left panel, \citealt{gravity2018b, gravity2020a,gravity2020c,gravity2023b,wielgus2022}). The accretion flow in the Galactic Center thus is a variant
of the `MAD' flows ($B\sim80$--100 G, \citealt{yuan2014,
bower2018, dexter2020, ressler2020a}).

The hot gas density in the accretion zone around SgrA* is comparably
low, $n_{e}\sim$ a few $10^{3}\, \mathrm{cm}^{-3}$ at a few
$10^{3}\, R_{g}$ \citep{gillessen2019}, and
$n_{e}\sim$ a few $10^{6}\, \mathrm{cm}^{-3}$ at $\geq 10\, R_{g}$ \citep{marrone2007, quataert2004}. The accretion flow in the Galactic Center thus is
\emph{radiative inefficient} and hot, since the density is too
low to equilibrate the electron and ionic accretion fluids \citep{rees1984,
quataert2000, yuan2003, yuan2014}. These properties are consistent with strongly sub-Eddington accretion
($10^{-8\ldots-9}\, M_{\odot}\, \mathrm{yr}^{-1}$, \citealt{baganoff2003,blandford1999a,gillessen2019} and references
therein). If these motions can be interpreted as Keplerian circular
orbits of hot spots, the astrometric data probe the potential on 8--10
$R_{g}$ scales (\citealt{broderick2006}, but see \citealt{matsumoto2020}). Combining the strong magnetic field and
low density it is then tempting to conclude that the near-face-on
orientation of the accretion flow as found by GRAVITY and ALMA reflects
the angular momentum of the accretion flow at large distances \citep{ressler2018,ressler2020b}. And indeed the observed orientation of the flow
deduced from the infrared flares is consistent with the angular momentum
direction of the `clockwise' disk of O/WR stars at distances of 1--3''
from SgrA* \citep{gravity2023b}. The winds from these
stars currently dominate the accretion flow onto SgrA* \citep{ressler2020b}.

\subsection{The Event Horizon Telescope and the detection of the
`shadow' as predicted by GR}

The central panel of Fig.~\ref{fig:3b} shows an image of the compact radio
source SgrA* obtained with the `Event Horizon Telescope', an array of
seven telescopes across the globe, and measuring the 1.3-mm continuum
radiation of SgrA*
(\url{https://eventhorizontelescope.org/},
see the more detailed discussion in Fig.~\ref{fig:A1b}). The EHT is
the pinnacle (in terms of resolution and short wavelength coverage) of
the classical radio spatial interferometry Fourier inversion technique
(cf.\ \citealt{thompson2017}). Once the collected data are
calibrated and analyzed, the EHT can reconstruct the 1.3mm brightness
distribution with a resolution of better than 20 micro-arcsec \citep{eht2022a, eht2022b}. As expected from theory (Sect.~\ref{sec:1}), the image
shows a bright ring with a central dip of diameter $51.8\pm2.3$
micro-arcsec. Given the mass and distance of SgrA* measured with high
precision from the stellar orbits \citep{eht2022a,eht2022b,gravity2023b}, GR of a near-face-on accretion zone of moderate to
low spin predicts a shadow of diameter $52\pm0.4$ micro-arcsec \citep{johannsen2013, johannsen2016b}, in excellent agreement with the EHT
measurement \citep{eht2022a, eht2022b} and a Kerr metric around a 4.3
million solar mass MBH.

The key finding thus is that the \emph{near-event horizon-scale
motions from GRAVITY and the size of the `shadow' of the EHT 1.3-mm
emission at 6--9\, $R_{g}$ are consistent with the GR `shadow'
computed from the prior of mass and distance determined from the stellar
orbits at a few $10^{3}\, R_{g}$.} This
indicates that any extended mass component within the peri-center
motions of the inner most stars is less than a few thousand solar masses \citep{gravity2022a,will2023}. Both the GRAVITY
motions and the EHT size thus constrain the compactness parameter
$\epsilon$ to be 0.4--0.6, fully consistent with a single MBH in GR (bottom
right inset).

The EHT has also detected the predicted shadow in the massive central
galaxy of the Virgo galaxy cluster, M87 (or Virgo A) \citep{akiyama2019}. The distance of M87 is 16.8 Mpc, 2000 times further away than
SgrA*. Since its mass of the central SMBH is 1500 larger than that of
SgrA*, $6.2$--$6.5\times10^{9}\, M_{\odot}$, the diameter of
its shadow is $42\pm3$ micro-arcsec, comparable to that of SgrA*, and
strengthening further the `shadow' interpretation. Detailed polarization
images of M87 with the EHT that the near-event horizon magnetic field
structure is poloidal and the region is magnetically dominated, as in
SgrA* \citep{akiyama2021a, akiyama2021b}.

\paragraph{Is there a relativistic jet emanating from SgrA*?}

Given the strong magnetic field in the accretion zone, and the evidence
for a low value of the gas to magnetic pressure, $p(\rm gas)/p(B)\leq1$ \citep{gravity2018b, gravity2020a, gravity2021a, gravity2023b, wielgus2022}, one
might expect a prominent relativistic radio jet from SgrA* \citep{falcke2000b}. Moreover, if the BH spin were substantial, one would
expect that the \cite{blandford1977} mechanism might be effective
in accelerating the nuclear spin-driven outflow. Yet, so far, no radio
jet feature has been detected, even in relatively high frequency
86\,GHz-VLBI maps with exquisite sensitivity \citep{issaoun2019}, where
foreground electron scattering should be less effective in smearing out
the jet feature. It is possible that the relative face-on orientation of
the central accretion zone would expect the spin to be along the line of
sight and thus hard to detect. Alternatively, the SgrA* spin might be
low, on the grounds that there is no evidence for strong accretion
events in the last few Myrs that could have spun up the MBH \citep{genzel2010}.

\paragraph{Lack of hard surface?}

The right panel of Fig.~\ref{fig:3b} shows the observed radio to X-ray
spectral energy distribution of SgrA*, with red labels pointing out the
origin of the data (see \citealt{genzel2010, gravity2020c, falcke2000b, melia2001}). The radio to NIR emission exhibits
substantial linear polarization, pointing to non-thermal synchrotron
emission from the hot accretion flow, as well as possible jet driven
outflows. The K- to mid-IR emission is characterized by a
semi-continuously variable, linearly polarized source, with a near
log-normal flux distribution over 1.5 decades \citep{genzel2003,eckart2006a,eckart2006b,dodds2009,dodds2010,dodds2011,do2009,witzel2012, witzel2018,gravity2020c}. The X-ray
emission is also highly variable over two orders of magnitude and could
come from synchrotron emission as well
\citep{dodds2010,ponti2017,gravity2021b}, or, perhaps less likely, from
Compton up-scattering of long wavelength photons (\citealt{dodds2010,witzel2021}, see also \citealt{genzel2010, cardoso2019}). \cite{broderick2009} have used the relative weakness of the steady infrared
emission from the center of SgrA* ($<$ a few percent of the
total luminosity of a few $10^{36}\, \mathrm{erg\
s}^{-1}$) as an argument in favor of the existence of an
event horizon (but \citealt{abramowicz2002,carballo2023}). The basic argument is as
follows. Assume matter were accreting onto a hypothetical hard surface outside the
gravitational radius, but within the upper limit of $\sim10
R_{g}$ set by the VLBI images. When this accreting matter
hits the surface, it will shock, thermalize, and emit all its remaining
energy as black-body radiation of a few $10^{3}$ K in the
IR range. Such a thermal component is not observed in the steady
spectral energy distribution of SgrA* \citep{genzel2010, gravity2020c}, setting a stringent upper limit on the mass accretion rate. In
practice this limit is so low that even the low level of observed
quiescent radio/submm nonthermal emission requires an assumed radiative
efﬁciency of nearly 100\%. This can be ruled out, which then leads to
the conclusion that the central object cannot have a hard surface but
rather must have an event horizon ($\epsilon \ll 1$). The caveat is
that this consideration does not include gravitational light bending.
\cite*{lu2017} make a similar argument on the statistical
lack of bright extragalactic tidal disruption events in the Pan-STARRS1
survey.

\section{\textit{Allegretto:} GRAVITY measurements of BH masses in distant AGN and
quasars}\label{sec:4}

We discussed in Sect.~\ref{sec:2} the discovery of quasars 60 years ago,
interpreted to be (S)MBHs ($10^{7}$ to
$10^{10}\, M_{\odot})$ at large distances, and
accreting gas at large rates. Large samples of these active galactic
nuclei (AGN, of which quasars are the tip of the ice-berg) are now
available, all the way back to less than 1 Gyr after the Big Bang, in
the Early Universe. In the local Universe the mass of the central MBH
and the mass of the central stellar centroid/bulge are about 0.2--0.7\%
of their host galaxy (Fig.~\ref{fig:1} right panel). This correlation suggests that the evolution and
growth of the two components are correlated on cosmic time scales
(\citealt{vestergaard2008, vestergaard2009, alexander2012, kormendy2013, mcconnell2013, saglia2016}, and references therein).

How can we test more quantitatively the correctness of the (S)MBH
paradigm and the (S)MBH-galaxy co-evolution? `Type I' AGNs and quasars
typically show $>$ few $10^{3}$ km/s line
broadening of atomic emission lines by high-velocity motion of gas near
the center \citep{netzer2013, netzer2015}. If the line widths are due to virialized
motions caused by the central mass, and the size of regions for which
broad emission lines are observed (`\emph{broad-line regions,
BLR}') it then becomes possible to measure the masses of the central
(non-stellar) masses in individual AGNs.

One way to estimate the BLR sizes comes from measuring the delay in
light travel time between the variable brightness of the accretion
disk~continuum and the emission lines, a method known as
`\emph{reverberation mapping}` \citep{blandford1982}. This
method has been applied routinely to nearby AGNs (e.g. \citealt{kaspi2000, bentz2009, bentz2010, bentz2013, peterson2014}) but is somewhat
limited because of the necessary underlying assumptions on the source
structure and geometry.

Moreover, ground-based reverberation mapping cannot be easily applied to
the very massive, large quasars because of the long time scales
involved. Until recently direct imaging of such BLRs has not been
possible because of their large distances and resulting small angular
size (less than 100 micro-arcsec). For the most distant SMBHs the
approach has been to calibrate the relationship between BLR size
(inferred from reverberation) and optical luminosity, and then apply the
same relationship to higher $z$ SMBH where only line widths and
luminosities were available \cite{vestergaard2008, vestergaard2009}.

\begin{figure}[htbp!]
    \centering
    \includegraphics[width=1\linewidth]{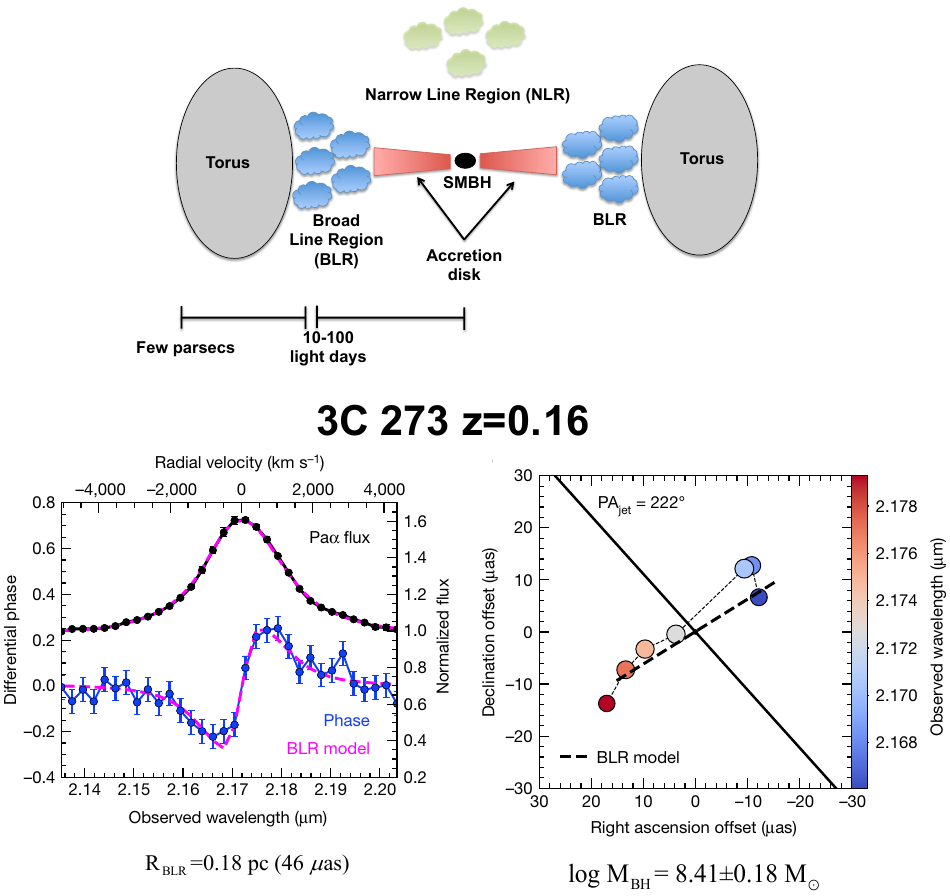}
    \caption{\textbf{Top:} Schematic of the structures around a luminous,
rapidly accreting extragalactic active galactic nucleus $(\dot{M} > 0.01 \times \dot{M}_{\max, \rm Eddignton})$,
with a super-massive $(>10^{8\ldots10}\,M_{\odot})$ BH at its center (e.g. \citealt{osmer2004, netzer2015}). The
SMBH is surrounded by a hot accretion disk. Generalizing our current
GRAVITY results (Figure 3, \citealt{gravity2018c, gravityplus2024}, and references therein) on its outer side are
self-gravitating ionized clouds, the BLR in virial equilibrium and
rotating around the SMBH. This central region in turn is surrounded by a
dusty molecular region (the ``Torus''), and ionized clouds on 100 pc-10
kpc scale (the narrow-line region, NLR). Image credit: Claudio Ricci. \textbf{Bottom:} GRAVITY
spectro-astrometry of the broad P$\alpha$ line in the $z=0.16$ Quasar 3C 273
\citep{schmidt1963}. The left panel shows the observed line profile and the
inferred interferometric phase gradient across the line, with a
measurement accuracy of about 1$\mu$as (500 AU, or 0.0024 pc at 550 Mpc
distance). The 2D spectro-astrometry model (bottom
middle panel) shows that this phase gradient extends over 50
micro-arcsec, approximately perpendicular to the direction of the known
radio jet (black line). The model yields a MBH mass of $2.6\times10^{8}\, M_{\odot}$, surrounded by a thick
rotating gas disk of 0.18 parsec (46$\mu$as) diameter (\citealt{gravity2018c}, cf.\ \citealt{netzer2020}).}
    \label{fig:4a}
\end{figure}

\begin{figure}[htbp!]
    \centering
    \includegraphics[width=1\linewidth]{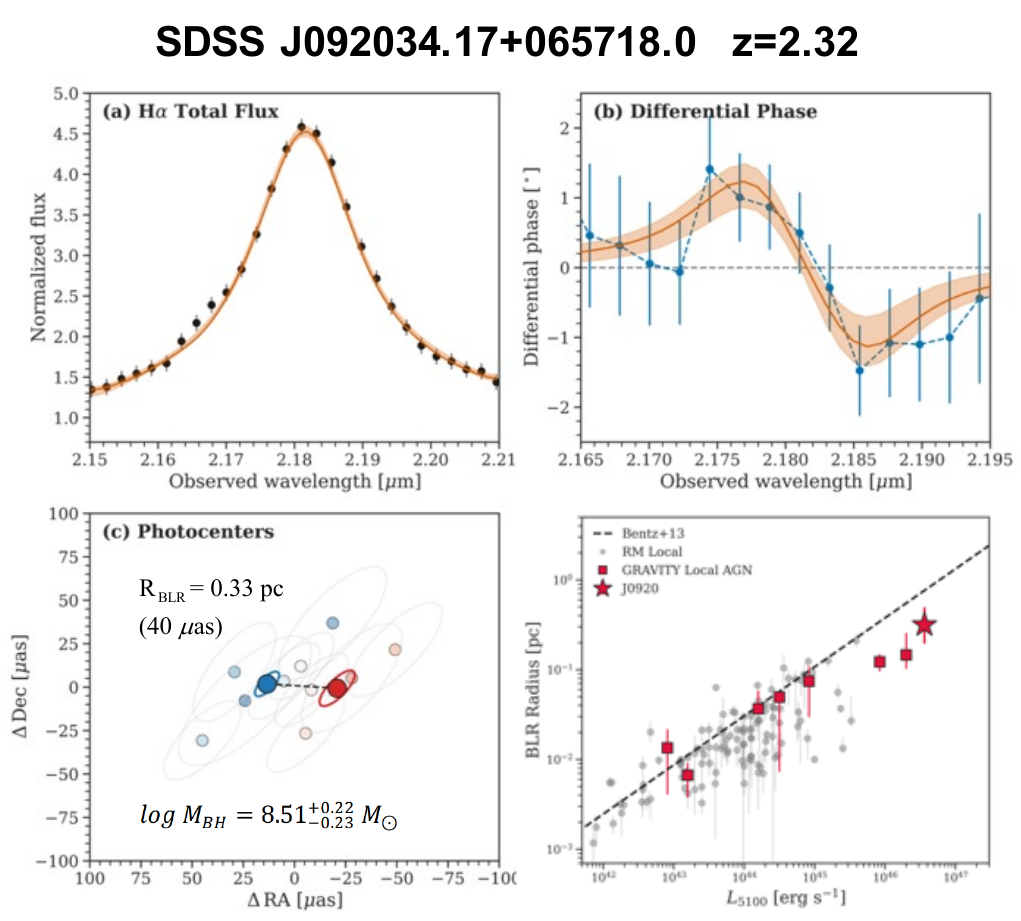}
    \caption{\textbf{Top left:} GRAVITY spectro-astrometry of the
broad H$\alpha$ line in the $z=2.32$ quasar J0920, with the new capabilities of
`GRAVITY-wide' to fringe-track on a bright star up to 25'' from the
phase center \citep{gravityplus2024}. The data and modeling
of this distant quasar (distance 17,700 Mpc, 2.89 Gyr after the Big
Bang) are qualitatively quite similar to those of the 33 times closer 3C
273, with a linear velocity gradient, indicating predominantly rotation
in a 0.33 pc diameter (40 micro-arcsec) thick disk rotating around a
$3.2\times10^{8}\,M_{\odot}$ central mass (top right and
bottom left). In contrast to moderately luminous local AGN, 3C~273
and J0920, both optically very luminous (i.e. high Eddington ratio
accretion) MBHs, seem to have a smaller broad-line region, and contain a
smaller fraction of the overall baryonic mass of the galaxy than
predicted by scaling relations (bottom right).}
    \label{fig:4b}
\end{figure}

The GRAVITY interferometric beam combiner of the $4\times8$-m telescopes of the
ESO-VLT has changed this situation (Fig.~\ref{fig:A1a}, \citealt{gravity2017, gravityplus2022}). \cite{gravity2018c} were able to resolve the BLR of the famous quasar
3C~273 ($z=0.16$, 550 Mpc distance, cf.\ \citealt{schmidt1963}) at sub-parsec level
with interferometric spectro-astrometry. Figure~\ref{fig:4b} shows
the results. More recently, after incorporating new optics allowing much
larger offsets between phase reference star and science object, it has
now become possible to apply the same direct technique to faint, distant
quasars, such as the quasar J0920 ($z=2.32$, 17,700 Mpc distance, \citealt{gravityplus2024}; Fig.~\ref{fig:4a}). At this time, there are about a
dozen GRAVITY(+) measurements of BLRs (bottom right of
Fig.~\ref{fig:4b}).

Looking ahead at the near-future, the enhanced capabilities of
GRAVITY+ will enable measuring super-massive BH masses
and their evolution across the entire cosmological evolution of
galaxies, and answer the fundamental question whether galaxies, and MBHs
grew in lockstep, or whether one of them grew faster and earlier. The
high-$z$ detections with the new interferometric technique of optically
very luminous (i.e. high Eddington ratio accretion) (S)MBHs, may have
systematically smaller broad-line regions, and thus contain a smaller
(S)MBH mass, than lower Eddington rate AGNs measured in the local
Universe with the reverberation technique, and with the
velocity-luminosity relations at higher $z$ (bottom right of Fig.~\ref{fig:4b}). Even small shifts as seen here could have big impacts on our
understanding of (S)MBH growth as these relations are used to measure
masses out to $z=10$ in usually high luminosity/Eddington quasars and
discriminate between black hole seed models.

\clearpage
\section{\textit{Allegro Molto:} Experimental evidence for stellar BHs from
gravitational waves}\label{sec:5}

\begin{figure}[ht!]
    \centering
    \includegraphics[width=1\linewidth]{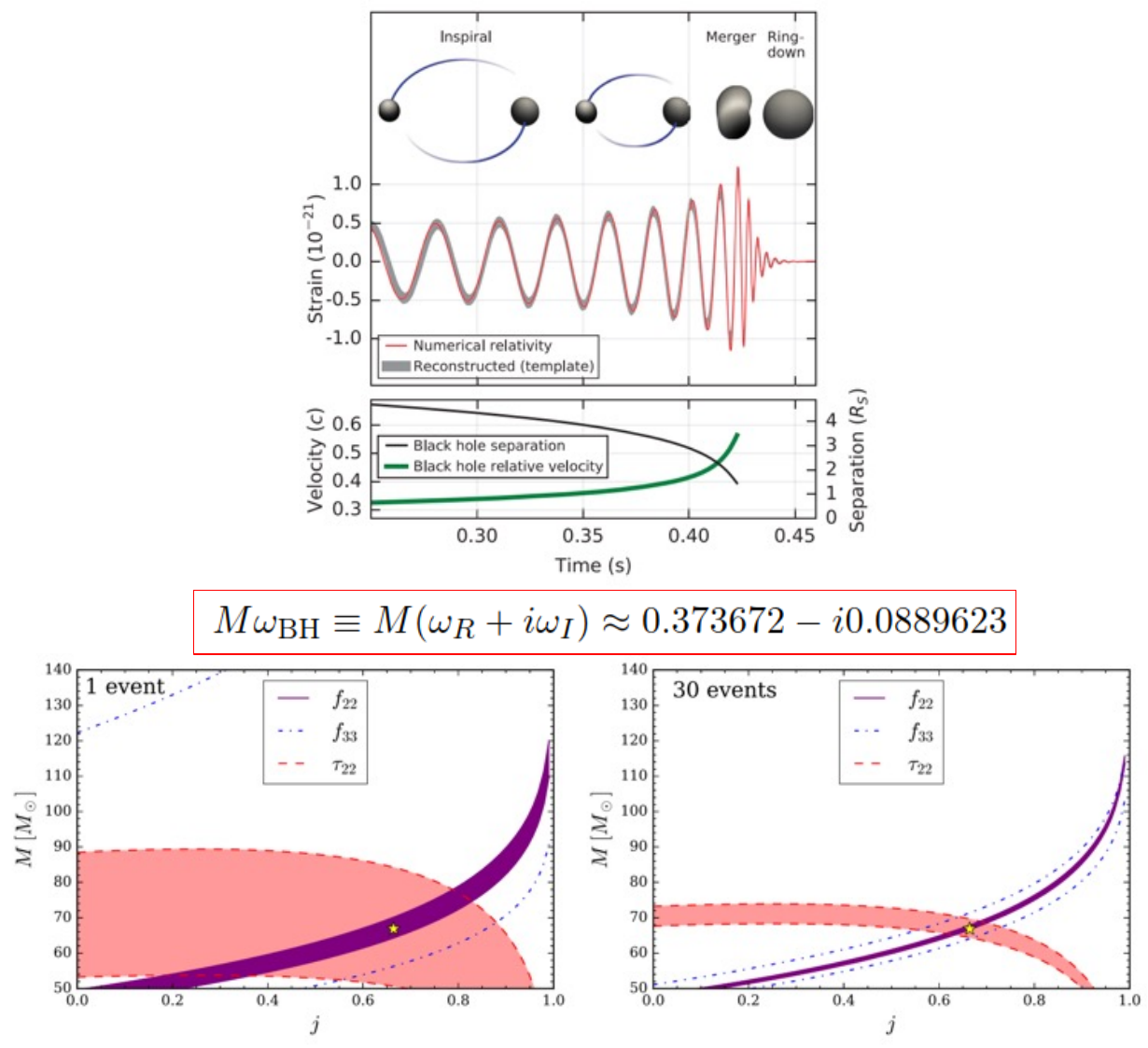}
    \caption{Top: Inspiral, merger/plunge-in and ring-down of a SBH
binary \citep{abbott2016a}. Bottom: `Spectroscopy' of a SBH inspiral.
Mid and right: The Kerr metric gives a unique relation-ship between
mass, and the orbital $l=2$ mode frequency $\omega_{R}$ near
the plunge in, and the imaginary frequency $\omega_i$
expressing the decay time of this mode, $t_d \sim1/\omega_i$ \citep{cardoso2019}. Given the short
duration of a $2\times30\, M_{\odot}$ inspiral, a single inspiral like
GW150914 does not yield enough SNR to determine these frequencies with
sufficient accuracy with the current aLIGO sensitivity, and a stacking
of about 30 such inspirals would be required (simulation by \citealt{brito2018}). The small yellow star is the true input value
injected into the simulation.}
    \label{fig:5}
\end{figure}

If two compact stellar masses orbit each other in a tight bound orbit,
they lose energy through the emission of gravitational waves. As a
result, the orbital semi-major axis shrinks. The inspiral rate is low
initially at large orbital radius, but then orbital speed, gravitational
wave strain and gravitational wave amplitude increase as the inspiral
proceeds. Once the binary has shrunk to the innermost stable orbit,
plunge in to a SBH of the combined mass (minus the mass-energy lost due
to gravitational wave emission) happens on a dynamical time scale of O
(10milli-sec) or 200 Hz frequency (left panel of Fig.~\ref{fig:5}). The
gravitational strains that need to be detected for successful of binary
inspirals lead to pico-meter amplitudes, which can be detected with
second and third generation laser-powered Michelson interferometers
(\url{https://www.ligo.caltech.edu/},
\url{https://gcn.nasa.gov/missions/lvk}).

\begin{figure}[ht]
    \centering
    \includegraphics[width=1\linewidth]{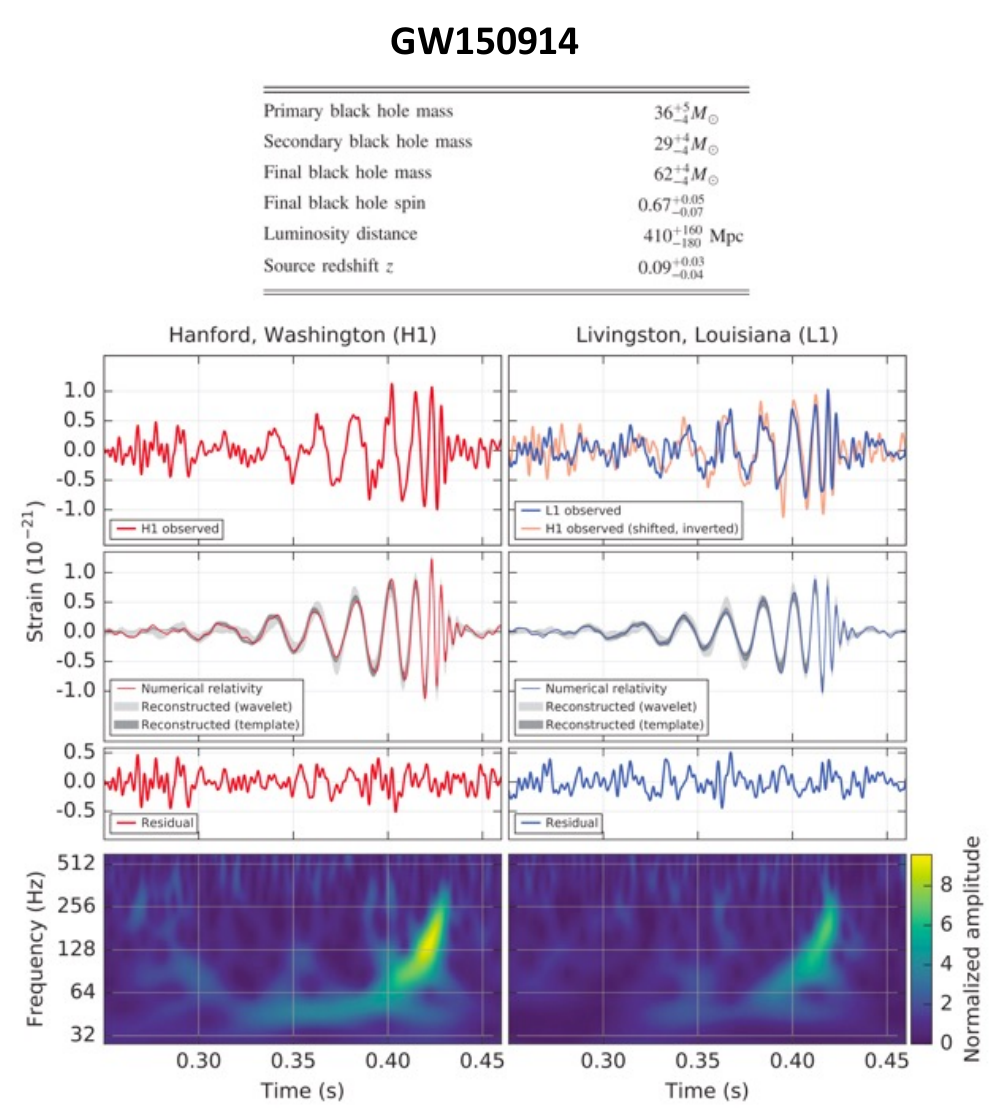}
    \caption{ Data of the first BH-binary
inspiral, GW150914, as seen by the Hanford and Livingston antennas of
aLIGO, and the derived source properties |\cite{abbott2016a, abbott2016b}.
}
    \label{fig:6a}
\end{figure}

\begin{figure}[ht]
    \centering
    \includegraphics[width=1\linewidth]{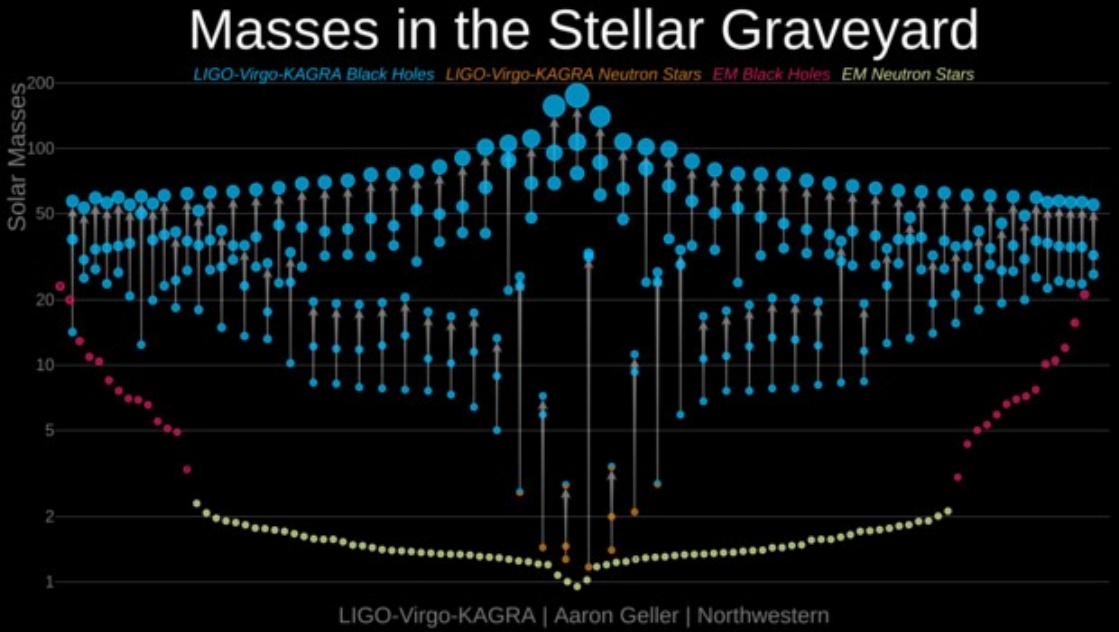}
    \caption{Current status of the SBH-SBH and SBH-NS inspirals observed
by aLIGO and aLIGO+Virgo+KAGRA after GWTC-3 (Fig.~\ref{fig:A1c}, right panel, \citealt{abbott2016a, abbott2016b, abbott2022}). Compilation of the inferred inspiral masses of all events seen at the end
of the GWTC-3 run with LIGO, Virgo and KAGRA \cite{abbott2023}. Image
courtesy of \url{https://www.ligo.caltech.edu/image/ligo20211107a}.
}
    \label{fig:6b}
\end{figure}

The first spectacular detection was the gravitational-wave pattern from
GW150914, with the two gravitational-wave antennas of LIGO in the USA
\citep{abbott2016a, abbott2016b}. The two initial (S)BHs had masses of about
$35\, M_{\odot}$. Since 2015 and up to the end of the GWTC-3 run,
aLIGO, strengthened with the Virgo antenna in Europe, and the KAGRA
antenna in Japan, have detected 35 bona-fide mergers. These turn out to
be SBH-SBH and SBH-NS mergers, but no NS-NS mergers, as well another
$\sim$50 or so candidate detections \citep{abbott2022}. This
impressive harvest is shown in the in Fig.~\ref{fig:6b}. The data
provide very strong, and arguably conclusive evidence for the existence
of SBHs.

Starting already with GW150914 (Fig.~\ref{fig:6a}) it perhaps came
as a surprise that the SBH masses were larger than expected. According
to standard stellar evolution models, the typical SBH in the local
Universe is expected to have a mass of $\sim10\,M_{\odot}$. While a number of such lower mass inspirals have by
now been seen, the relative fraction of $\sim2\times 25$--$30\,
M_{\odot}$ objects is fairly, if not uncomfortably, large. Even
more surprising is the case of GW190521, with two SBHs of 85 and 66
$M_{\odot}$ combining to an `intermediate mass BH' (IMBH) of 142
$M_{\odot}$ \citep{abbott2020}. The two initial SBHs of the
binary are both in, or near the so called `pulsational pair instability
gap', where the creation of particle-antiparticle pairs during the
supernova explosion there prevents a stable SBH mass. While the large
fraction of combined mass $\sim50\, M_{\odot}$ SBH end
states may be explainable by the instrumental sensitivity bias of the
current interferometers preferentially selecting larger mass, larger
amplitude systems, the case of GW190521 is truly fascinating. Do
repeated mergers in dense star clusters explain the large masses, or is
there a new, yet unknown channel of massive SBH and IMBH creation?

\paragraph{Black-hole spectroscopy}

It is clear that the largest information about the space-time and
significant tests of the no-hair theorem can in principle obtained in
the last, near-event horizon inspirals, before plunge in. In technical
terms this `BH spectroscopy' requires accurate determination of the
near-photon orbit, orbital frequency and its decay time (\citealt{brito2018, berti2018, cardoso2019}; Fig.~\ref{fig:5}).
The Kerr metric specifies a precise relation between the $l=2$ mode
`normal' frequency (at the photon orbit), and the decay time (Fig.~\ref{fig:5}).
The frequency analysis of the GW150914 ring down (the strongest and highest quality inspiral as of today) is consistent with Fig.~\ref{fig:5} in the $l=2$ mode, and thus with the Kerr metric \citep{abbott2016b}. However, \cite[bottom panels of Fig.~\ref{fig:5}]{brito2018} have shown that a full test of the no-hair theorem requires also the determination of the frequencies and decay times of the $l>2$ modes, which is not yet possible with one inspiral. They estimate that about 30 inspirals need to be combined to achieve that test.
However, there is very good hope
that the ground-based interferometer of the next generation and the
space interferometer LISA will make dramatic improvements.

\section{\textit{Rondo:} Dark matter cusps}\label{sec:6}

\cite*{viollier1993} and \cite*{munyaneza1999} proposed that the dark matter concentrations in galactic
nuclei, including QSOs, may not be Kerr BHs, but very compact
`\emph{fermion balls'} (for example made of hypothetical,
massive neutrinos) supported by degeneracy pressure. The size of a
fermion ball, and the maximum stable `Chandrasekhar' mass (or its
relativistic analog, the `Oppenheimer--Volkoff' mass
$M_{\rm OV}$, \citealt{oppenheimer1939}), increase the
lighter the fermion's mass. In this scenario the largest observed
central masses in elliptical galaxies, a few $10^{10}\,M_{\odot}$,
would approach the Oppenheimer--Volkoff mass, resulting in an upper limit
to the mass of the constituent fermions to about
$13\, \mathrm{keV}/c^{2}$, again with larger BH masses requiring
smaller neutrino masses. To still `fit' within a given peri-center
distance the neutrinos would have to have a mass of \citep*{munyaneza1999}
\begin{equation}
    m_{\rm f} \sim 70\, \mathrm{keV\, c}^{-2} \left(\frac{R_{\rm peri}}{10 \, \text{light hours}}\right)^{-3/8} \left(\frac{g}{2}\right)^{-1/4} \left(\frac{M_{\rm OV}}{4.4 \times 10^6 \, M_{\odot}}\right)^{-1/8}
    \label{eq:3}
\end{equation}
where $g$ is the spin degeneracy factor of the fermion.

\begin{figure}[ht]
    \centering
    \includegraphics[width=1.0\linewidth]{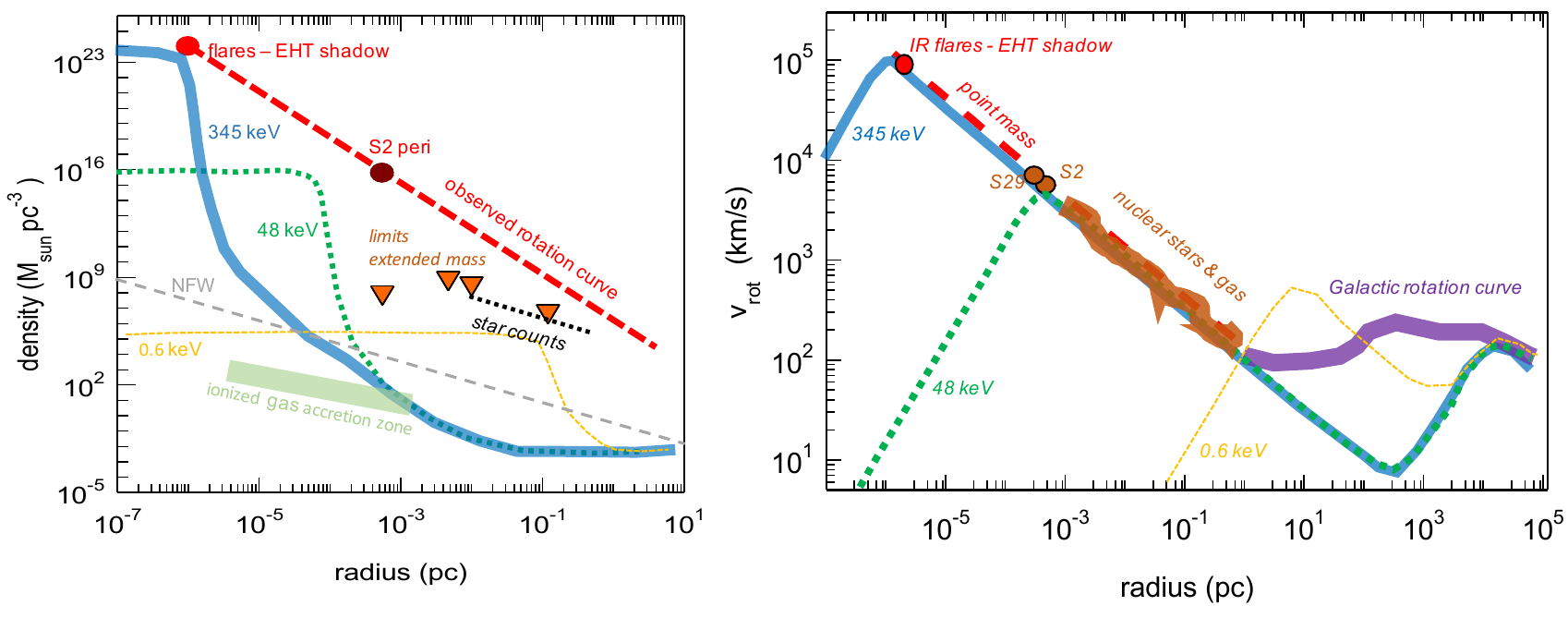}
    \caption{Constraints on the contribution of various mass components of
baryonic and dark fermionic matter (fermion ball) to the central mass
density (left) and rotation velocity (right) at different radii $R$
(adapted from \citealt{arguelles2019, arguelles2023}).}
    \label{fig:7}
\end{figure}

\cite*{ruffini2015} and \cite{arguelles2019} have pointed out that a self-gravitating equilibrium distribution
of massive neutral fermions of given degeneracy g exhibits a segregation
into three physical regimes. There is an inner core of almost constant
density governed by degenerate quantum statistics. Surrounding this core
is an intermediate region with a sharply decreasing density
distribution, which in turn is surrounded by an extended plateau.
Finally, there is an outer, asymptotic halo where density scales as
$\rho \propto R^{-2}$, i.e., a classical Boltzmann regime
(Fig.~\ref{fig:7}, adapted from \citealt{arguelles2019, arguelles2023}. The \cite{ruffini2015} model
unifies dark matter on large, intergalactic and circum-galactic scales,
with the evidence of central massive concentrations discussed in
Sects.~\ref{sec:2}--\ref{sec:5}. The Boltzmann distribution would explain the ﬂat rotation
curves on $\gg$10 kpc scales, while baryons in form
of gas and stars would explain the mid-scales (parsec to 10+ kpc).
Figure~\ref{fig:7} shows the resulting density distributions and rotation curve
distributions as a function of radius, for three different `fermions',
of $m_{\rm f} =$\, 0.6, 48 and 345 keV. As pointed out
above, for a given fermion mass $m_{\rm f}$ in Eq.~\eqref{eq:3},
$M_{\rm OV}$ gives the critical stable mass, above which
gravity will take over and the object collapses (to a (S)MBH). Likewise,
if there is substantial baryonic accretion onto such a dark matter core,
the central mass concentration would also collapse to a (S)MBH
 \citep{arguelles2024}.

Figure~\ref{fig:7} summarizes the current constraints on the fermion ball, or
fermion-dark matter model in the Galactic Center. The outer flat
rotation curve of the Galaxy would be best matched by moderate mass,
10--50 keV fermions \citep{arguelles2019, arguelles2023, becerra-vergara2021}. The nuclear stellar and gas velocities \citep{gravity2022a} are in principle also consistent with a 40+ keV dark matter core
\citep{becerra-vergara2020, becerra-vergara2021, arguelles2023}. However, the
excellent agreement of the measured prograde precession of S2 at
$R_{\rm peri}=5.7\times10^{-4}\, \mathrm{pc}$ with that
expected for a MBH \cite{gravity2020a, gravity2022a} is barely consistent
with such a lower mass dark matter particle. There is also a modest
inconsistency with the density of ionized baryonic mass estimated at a
few $10^{3}\,R_{g}$ \citep{gillessen2019}.

If the infrared flare motions ($\sim0.28 c$ on the sky at 8--9
$R_{g}$, \citealt{gravity2018b, gravity2023b}) represent
the Keplerian orbital speed, and the EHT shadow diameter \cite{eht2022a, eht2022b} is a measure of the light bending by a central
mass, then the implied fermion mass would have to be $\sim$500
keV, obviously very close to the MBH solution. Such a large fermion mass
is disfavored for the favored dark matter agent on large scales \cite{arguelles2023}.
It would also be inconsistent with the mass
densities inferred from the BLR in AGN and quasars (see Sect.~\ref{sec:4}).

Another proposed non-BH configuration is the `\emph{boson
star}' scenario advanced by \cite{torres2000}. A wide range of boson
star masses can theoretically be imagined, including ones with masses of
(S)MBHs, depending on the assumptions about the specific boson particle
masses and their self-interactions. Since such an object consists of
weakly interacting particles it is unclear how it may have formed. The
size of a boson star is only a few times $R_{S}$ of the same
mass (S)MBH, and is highly relativistic. It is clear from Fig.~\ref{fig:7} that
even the astrometric observations of the IR flares are not sufficient to
distinguish a boson star from a MBH in the Galactic Center \citep{rosa2022}. The motion
of test-particles crossing the boson star would allow to distinguish
boson stars from BHs \citep{zhang2022}, but are currently far out of
reach. The EHT imagery is more promising. Ray-tracing simulations by
\cite{vincent2016} and \cite{olivares2020} suggest that the
mm-appearance of an accreting boson star can be close to that of a MBH,
although `ideal' cases with a large shadow as seen in the Galactic
Center, or M87 \cite{eht2022a, eht2022b, akiyama2019} are
rare. More detailed EHT observations are likely to separate in more
detail the time variable, from the stationary parts of the shadow, thus
helping to distinguish between the MBH and boson star cases.

Finally, Occam's razor strongly disfavors the boson star interpretation
for rapidly accreting AGNs/quasars. A boson star obviously is unstable
to collapse to a BH if it experiences substantial baryonic accretion.
While this is not the case in the Galactic Center currently, such high
Eddington ratio events almost certainly happened in the evolution of
most, and perhaps all, (S)MBHs.

Classical (particle-like) dark matter models predict that dark matter will cluster around the massive black holes in galaxy centers \citep{gondolo1999, sadeghian2013}. In principle, such dark matter spikes could have high enough densities to affect the motions of objects around the MBH \citep{zakharov2007,lacroix2018}. Yet, the detection of dark matter in that way might be impossible, due to the expected background population of dark objects such as white dwarfs, neutron stars and stellar black holes, that should arrange into a spiky distribution centered on the MBH \citep{merritt2010b, antonini2014, linial2022}, and that simply might outweigh the dark matter particles. Current limits from GRAVITY observations stand at around $4000\,M_\odot$ \citep{gravity2022a}.

\section{\textit{Coda Fortissimo:} Future expectations for studying astrophysical
BHs}\label{sec:7}

A century after the publication of Einstein's field equations and
Schwarzschild's first solution \citep{einstein1916, schwarzschild1916},
sixty years after the Kerr/Newman (1963-65) solutions, and the discovery
of X-ray binaries \citep{giacconi1962} and quasars \citep{schmidt1963}, BHs
have come from theoretical speculation to near experimental certainty.
The recent progress in electromagnetic and gravitational wave studies
has been truly remarkable, the top experiments (VLT-GRAVITY-Keck, EHT,
aLIGO-Virgo-KAGRA) have been a tour de force of experimental physics,
and the experimental work and scientific results have been recognized
with two Nobel Prizes, and several Balzan, Shaw, Gruber and Breakthrough
Prizes so far.

Yet let us stay realistic and humble. Measured on the expectation level
of critical theoretical colleagues, and certainly on the ultimate
requirement for establishing `scientific truth', \emph{the
evidence we currently have is impressive on all accounts but not (yet
fully) conclusive}. We have constrained the $\epsilon$-parameter to a few
tenths, leaving in principle open the possibility that the objects we
have been studying are not BHs after all but speculative ECOs, such as
fermion or boson stars, gravastars, wormholes etc. (see Sect.~\ref{sec:6}, and
the detailed review of \citealt{cardoso2019}). It is also possible that GR is not the correct description of space time close to the event horizon (cf.\ \citealt{cardoso2019} and references therein).

\begin{figure}[htb!]
    \centering
    \includegraphics[width=1\linewidth]{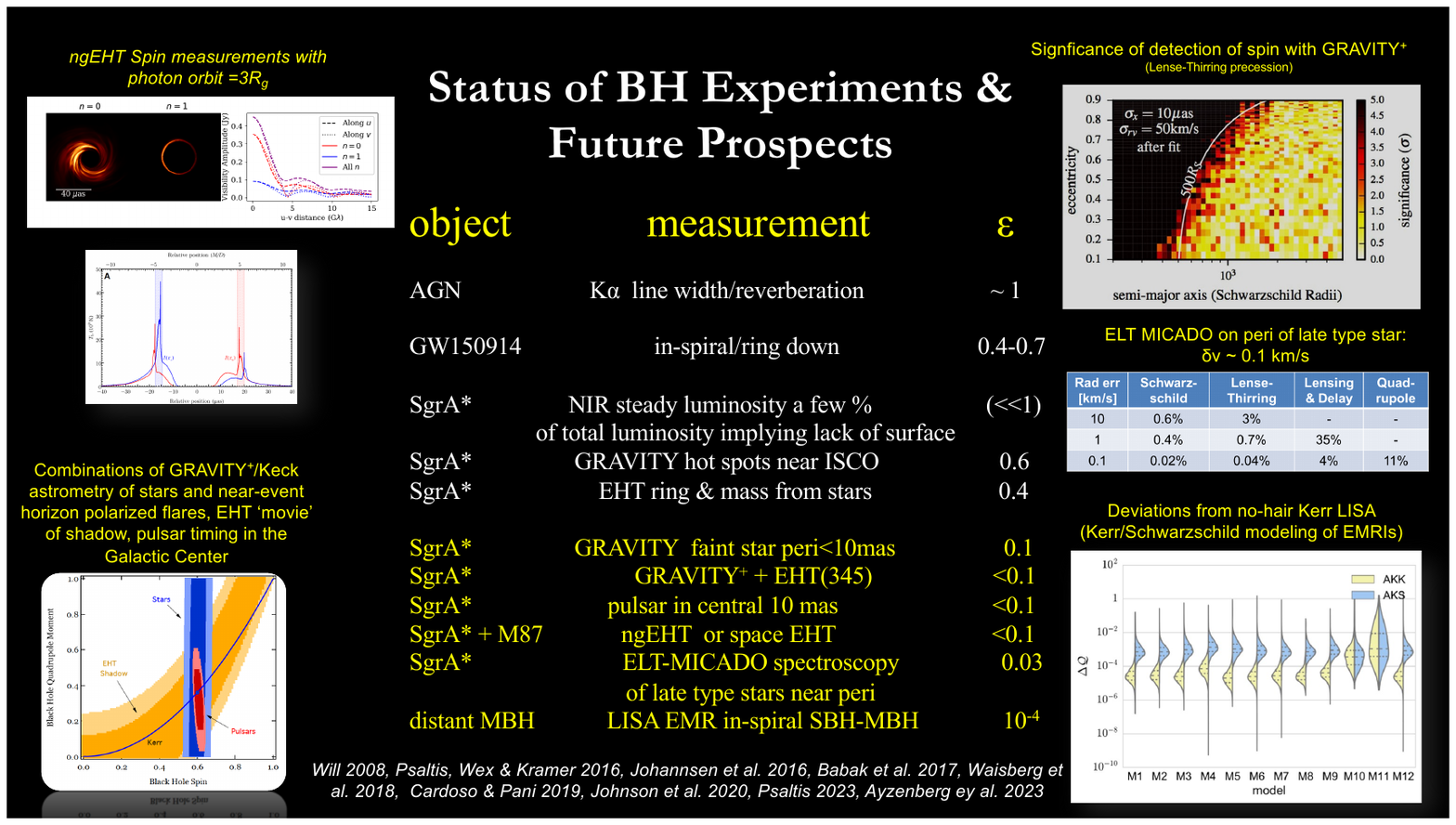}
    \caption{Current status and future improvements in the quality
of experimental studies of the BH paradigm. The central table lists the
constraints (here in the compactness parameter $\epsilon$, where $\epsilon=0$ is a Kerr
BH) achieved so far by the different techniques discussed in the text
are in white color, while the expected further improvements in the
future are in yellow. Current state of the art sets limits in $\epsilon$ of a few
tenths. The very faint stationary near-infrared emission of SgrA* can in
principle be interpreted as a strong evidence for the absence of a
surface of the source, and thus in favor of an event horizon. However,
this argument relies on the emission be isotropic and not strongly
affected by gravitational lensing. Detection with
GRAVITY+ of the Lense--Thirring precession of a star
with $R_{\rm peri}< 10 \,\mu{\rm as}$ would
yield a spin determination of the MBH in the Galactic Center, and
together with other stars yield a limit of $\epsilon\sim0.1$.
Higher quality measurements of the photon-ring ($n\geq1$) in SgrA* with
ngEHT, or space VLBI (together with the priors from
GRAVITY+ would reach $\epsilon<0.1$. The same
limit could be reached with timing of a Galactic Center pulsar within an
arcsecond of SgrA*. Still better limits could then come from a
combination of all three techniques in the next ten years. Detailed 100
m/s spectroscopy with MICADO@ELT in the 2030s of a late type star in a
close peri-approach to SgrA* might achieve $\epsilon\sim0.03$.
Finally gravitational wave analysis of an inspiral of a stellar BH into
a MBH (EMRI) would reach $\epsilon\sim0.0001$ with LISA data in
2+ decades.}
    \label{fig:8}
\end{figure}

In Figure~\ref{fig:8} below we have listed where we currently stand, and where we 
might get to in the next decades, for proving the MBH paradigm in terms
of the compactness parameter, or equivalently the no-hair theorem test,
with the parameter $\epsilon=0$ for a Kerr hole.

It is clear that the best LIGO-Virgo-KAGRA SBH inspirals already set
$\epsilon$ significantly below unity, likewise so for the most recent
combination of stellar, flare and EHT-shadow data. If
GRAVITY+ can measure a high quality orbit of a star
with a peri-center distance 2--4 times smaller than S2, or if one can
determine the pulse timing of a sufficiently nearby pulsar, or the
combination of GRAVITY+ and EHT (including possible
upgrades in performance), the next step is $\epsilon \leq 0.1$, allowing spin
measurements \citep{waisberg2018} and no-hair theorem tests (\citealt{falcke2000a, will2008, psaltis2011, psaltis2016, johannsen2016a, johnson2020}, lower top panel
of Fig.~\ref{fig:4b}). The enormously greater sensitivity of MICADO at the
ESO-ELT could push the spectroscopic measurement precision in 5--10 years
to m/s level \citep{davies2021}. For a late-type star near peri-center
that could push to $\epsilon \sim0.03$. Future extensions of the VLTI to a kilometer-wide interferometric array could push the angular resolution to better than 100 micro-arcsecond and astrometry to sub-micro-arcsec precision, thereby opening up the observation of the scattering of S-stars by neutron stars and stellar black holes, and spatially resolving the accretion zone and flaring activity at IR wavelengths. An expanded
ground-based mm-/submm-VLBI network, such as the ngEHT \citep{ayzenberg2023} could test the space-time at the photon-orbit. The European ELT
and GRAVITY+ will also probe the formation and
evolution of the first SMBHs, such as the JWST source GN--z11 at $z=10.6$
\citep{maiolino2023, schneider2023}.

An important question is, how far the measurement accuracy of the small higher order PPN-terms can be pushed before perturbations with other stars and SBHs introduce ‘orbital chaos’ that diminishes or even wipes out the GR information and significant tests of the no-hair theorem. \cite{merritt2010} carried out a suite of simulations and concluded that in the dense region around the MBH the perturbations indeed are likely too strong to measure the Lense--Thirring precession. However, the work of \cite{merritt2010} only considered orbit averaged quantities and assumed an overly high density of SBHs in the central region, which the current limits on any extended mass around SgrA* now show to be far too large. Moreover, we now know that many of the S-stars are on highly elliptical orbits and spend little time in the high density region around the MBH. The perturbative ‘chaos’ thus is dominated by single star-star and star-SBH events, when the three are near peri-approach. \cite{portegies2023} have studied this ‘punctuated chaos’ with high quality N-body integrations of the known S-star orbits. When two S-stars come very close (a few tens of AU) during peri of 100-1000 AU, they indeed experience exponential growth of the orbital deviations in their separation in parameter space. Still the average exponential growth time, the Lyapunov time, is ~460 years. This is encouraging if indeed faint stars of $<$100 AU peri-distance occur that in principle allow measurements of the Lense--Thirring and quadrupole terms \citep{waisberg2018}.

On a time scale of twenty years, the gravitational-wave mission LISA of
ESA should deliver the ultimate test. This space interferometer with
three satellites forming a laser interferometer of 2.5 million km length
is sensitive to gravitational waves with 4 to 5 orders lower frequency,
and thus correspondingly larger masses \citep{amaro-seoane2023, colpi2024}. LISA will be able to observe the inspirals of MBHs across
the entire Universe \citep{barausse2020}. The inspiral of a SBH into
a MBH (an extreme mass ratio inspiral, EMRI: \citealt{amaro-seoane2017,
amaro-seoane2019}) should deliver enough SNR in the inspiral before plunge, to
obtain a high-quality measurements of the fundamental ($l=2$)
quasi-normal mode of the MBH at the photon orbit, and get to $\epsilon\sim10^{-4}\ldots10^{-3}$
\citep{buonanno2007, babak2017, cardoso2019}. This would
provide the ultimate culmination of this exciting journey, which Albert
Einstein started more than a century ago.

\bmhead{Acknowledgements}
The authors are grateful to the numerous
contributions from GRAVITY team members in Germany, France, and
Portugal, as well as the generous support from ESO and their staff,
including the entire Paranal Observatory. None of the relevant research
discussed in this review would have been possible without this fantastic
team spirit and the key technical and scientific inputs from all our
colleagues. We are grateful to Vitor Cardoso, Thibault Damour, Hagai Netzer, Diogo Ribeiro, Taro Shimizu, and our
Editor, P.T.P. Ho, for helpful comments on the text.

\clearpage
\appendix
\section{Appendix: Instrumental
techniques}\label{appendix-instrumental-techniques}

All three experimental approaches discussed in this review apply
variations of the interference of light in a two-beam Michelson
interferometer \citep{michelson1887}.

\begin{figure}[!ht]
    \centering
    \includegraphics[width=1\linewidth]{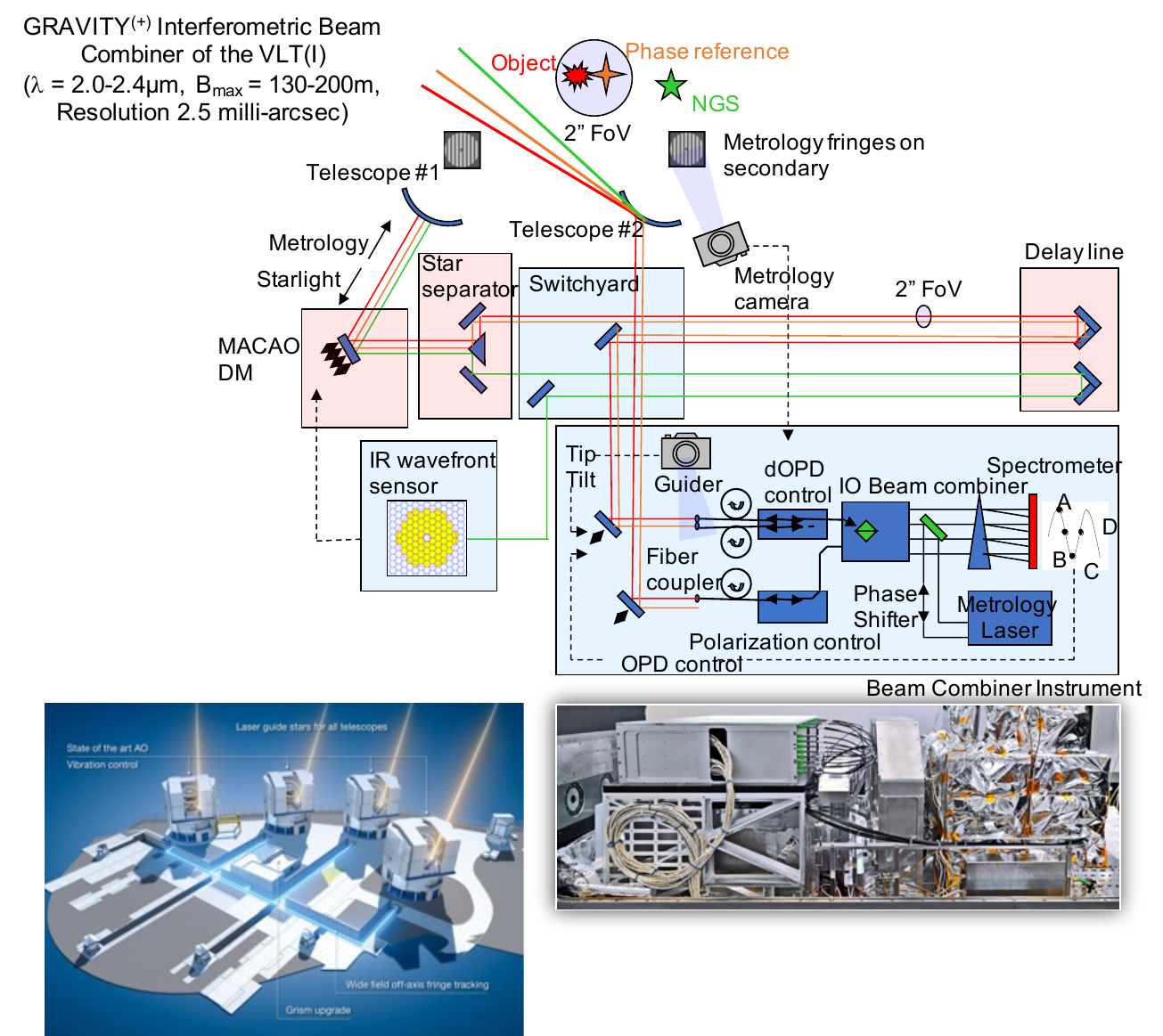}
    \caption{{\bf GRAVITY.} See text. Image credits: ESO/GRAVITY$^{(+)}$ collaborations.}
    \label{fig:A1a}
\end{figure}

\paragraph{GRAVITY.} Figure~\ref{fig:A1a} summarizes the
essentials of the stellar interferometry at 2$\mu$m (K-band) with the
GRAVITY(+) beam combiner at the ESO VLTI (\citealt{gravity2017, eisenhauer2023},
\url{https://www.mpe.mpg.de/ir/gravity},
\url{https://www.mpe.mpg.de/ir/gravityplus}). GRAVITY
coherently combines the light from the four 8m UT or 1.8m AT telescopes
(left bottom). Each telescope is equipped with adaptive optics to provide a
diffraction-limited beam, which is then transported through mirrored
delay lines to the cryogenic beam combiner instrument (right top). The
instrument provides two beam combiners, one for fringe tracking, the
other optimized for long exposure, high spectral resolution
interferometry of the science target. The optical path length within the
observatory is controlled via several laser metrology systems, delay
lines, and differential delay lines (top).

\begin{figure}[!ht]
    \centering
    \includegraphics[width=1\linewidth]{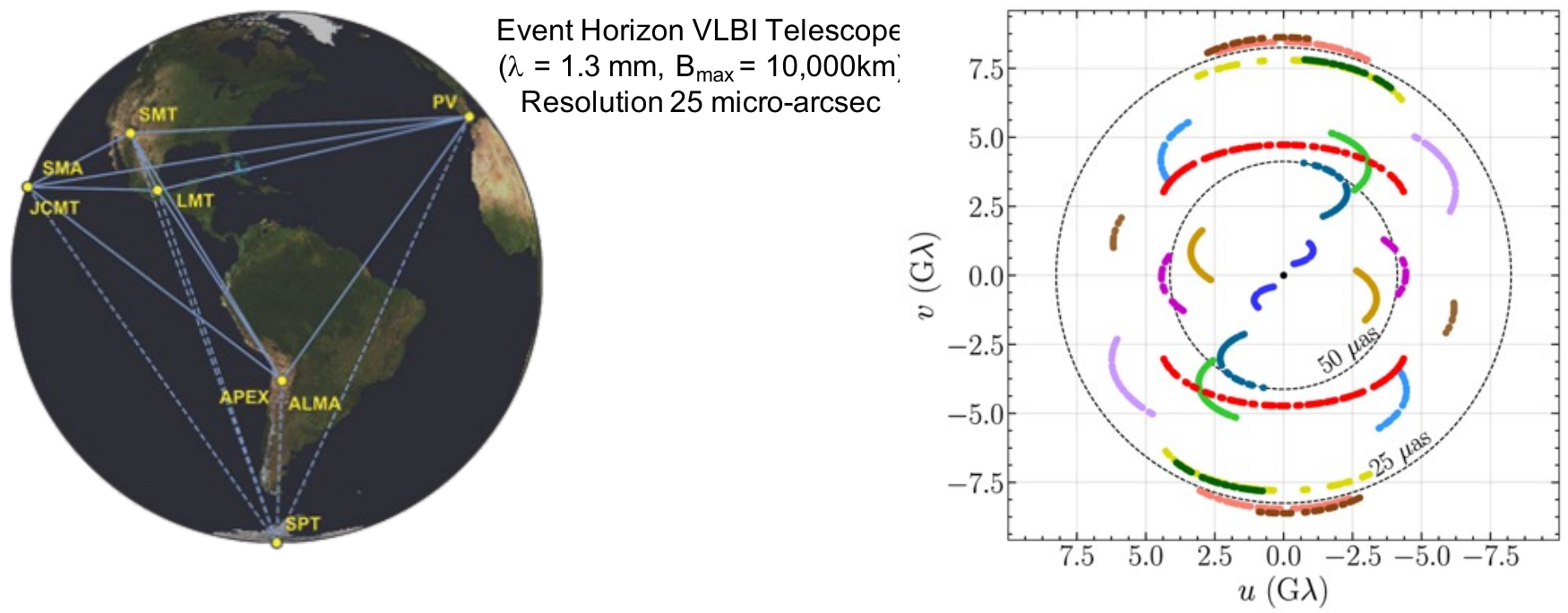}
    \caption{{\bf Event Horizon Telescope.} See text. Image credits:  EHT collaboration.
}
    \label{fig:A1b}
\end{figure}

\paragraph{Event Horizon Telescope.} The Event Horizon Telescope
(EHT) is a further development of the technique of intercontinental,
heterodyne stellar interferometry \citep*{thompson2017}
pushed to the highest microwave frequencies (230/345~GHz), at which the
Earth Atmosphere is still transparent. The Very Large Baseline
Interferometry (\citealt{eht2022a},
\url{https://eventhorizontelescope.org/},
\url{https://blackholecam.org/}) links radio telescopes across the
globe to create an Earth sized interferometer (left panel in
Fig.~\ref{fig:A1b}). Both techniques -- optical/IR and radio interferometery --
synthesize a virtual telescope of diameter
$B_{\max}$ (the maximum separation between two
telescopes in the array) with an angular resolution
$\lambda/B_{\max}$. They measure the Fourier
components of the image at the projected separation of the telescopes
(right).

\begin{figure}[!ht]
    \centering
    \includegraphics[width=1\linewidth]{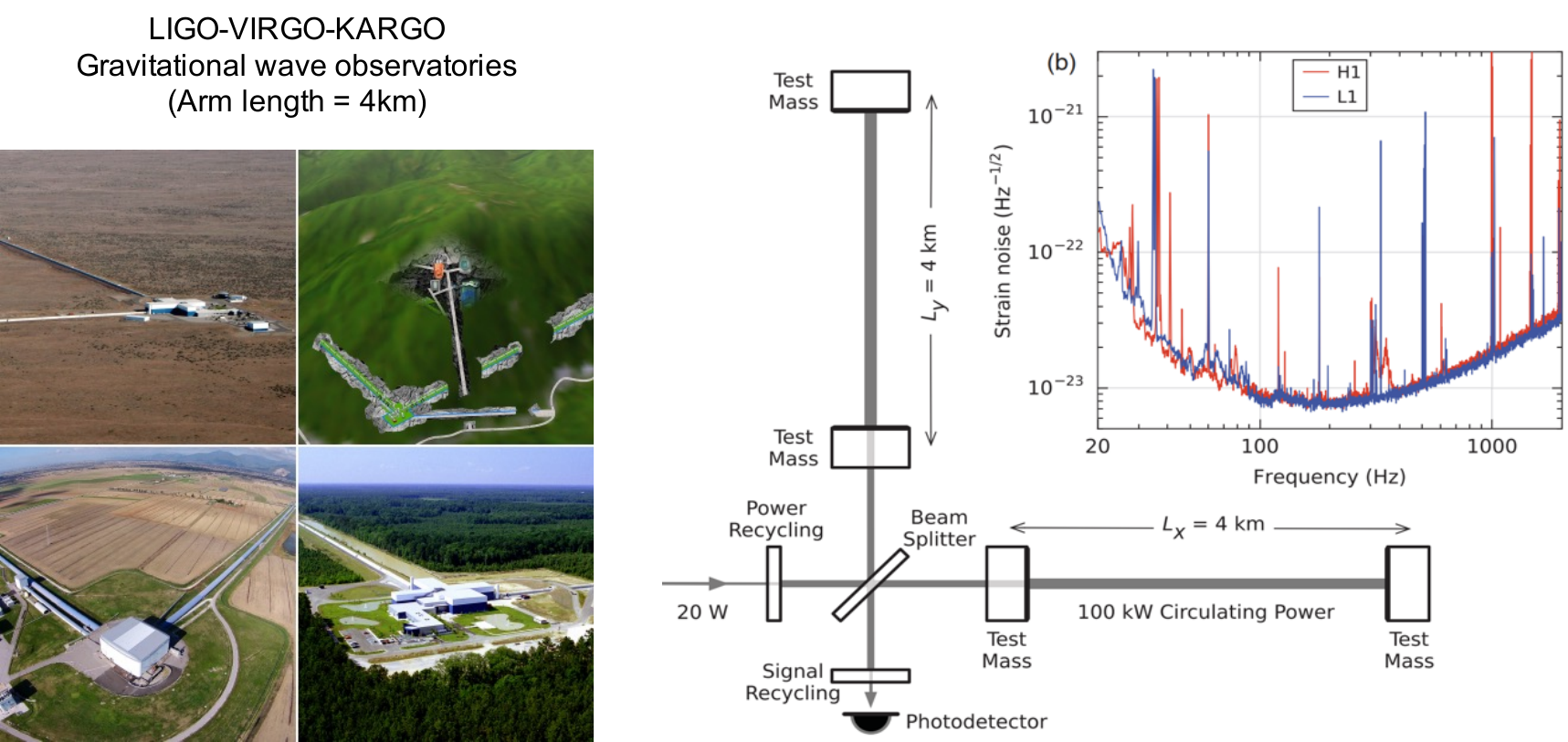}
    \caption{{\bf Ground-based gravitational-wave interferometry.} See text. Image credits: LIGO/Virgo/KAGRA collaborations.
}
    \label{fig:A1c}
\end{figure}

\paragraph{Ground-based gravitational-wave interferometry.} Figure~\ref{fig:A1c}
summarizes the essentials of ground-based gravitational
wave, laser interferometry, and in particular the LIGO, Virgo, KAGRA
gravitational-wave observatories (\citealt{abbott2016a},
\url{https://www.ligo.caltech.edu/},
\url{https://gcn.nasa.gov/missions/lvk}). The experiments
measure the small distortions of space-time when gravitational waves
pass through the two arms of the interferometer (right). In this case
one of the two arms will be stretched, while the other arm shortened.
Ground-based gravitational-wave detectors are most sensitive at
frequency of 20 to several hundred Hz (inset), which matches the orbital
frequencies of stellar black-hole mergers.

\clearpage

\phantomsection
\addcontentsline{toc}{section}{References}
\bibliography{genzel}

\end{document}